\documentclass[12pt]{article}
\usepackage{graphicx,psfrag,epsfig}
\usepackage{amssymb,amsmath,amscd,amsthm}
\usepackage{graphicx,psfrag,epsfig}

\usepackage{graphicx}
\usepackage[active]{srcltx}

\newtheorem{theorem}{Theorem}[section]

\newtheorem{lemma}[theorem]{Lemma}

\date{}

\begin{document}

\date{}
\title{On negative eigenvalues of low-dimensional Schr\"{o}dinger operators}
\author{
 S.
Molchanov\footnote{Dept of Mathematics, University of North
Carolina, Charlotte, NC 28223, smolchan@uncc.edu},
B. Vainberg\footnote{Dept of Mathematics, University of North
Carolina, Charlotte, NC 28223, brvainbe@uncc.edu ; corresponding author}
}
 \maketitle
\begin{abstract}
The paper concerns upper and lower estimates for the number of negative eigenvalues of one- and two-dimensional Schr\"{o}dinger operators and more general operators with the spectral dimensions $d\leq 2$. The classical Cwikel-Lieb-Rosenblum (CLR) upper estimates require the corresponding Markov process to be transient, and therefore the dimension to be greater than two. We obtain CLR estimates in low dimensions by transforming the underlying recurrent process into a transient one using partial annihilation. As a result, the estimates for the number of negative eigenvalues are not translation invariant and contain Bargmann type terms. We show that a classical form of CLR estimates can not be valid for operators with recurrent underlying Markov processes.  We provide estimates from below which prove that the obtained results are sharp. Lieb-Thirring estimates for the low-dimensional Schr\"{o}dinger operators are also studied.
\end{abstract}


\section{Introduction}

Let $N_0(V)=\#\{\lambda_j\leq 0\}$ be the number of non-positive eigenvalues of a Schr\"{o}dinger operator
\begin{equation} \label{sc}
H=-\Delta-V(x),~V\geq 0,
\end{equation}
on $R^d$ or $Z^d$. Everywhere below we assume that the potential is non-negative. The standard approach to Cwikel-Lieb-Rosenblum (CLR) estimates for $N_0(V)$ (see \cite{c}, \cite{L}-\cite{Lt1}, \cite{r}, \cite{rs}) requires the Markov process $x(t)$ which corresponds to the unperturbed operator $H_0=-\Delta$ to be transient (non-recurrent). The transience in the lattice case means that the expectation of the total time the process $x(t)$ spends in the initial point is finite. The latter is equivalent to the condition
\begin{equation}\label{rec}
\int_0^{\infty}p_0(t,x,x)dt<\infty,
\end{equation}
where $p_0(t,x,y)$ is the fundamental solution of the corresponding parabolic problem
$$
\frac{dp_0}{dt}=\Delta p_0,~t>0,~~ p_0(0,x,y)=\delta_y(x).
$$
In the continuous case, one needs to talk about the time spent in a neighborhood of the initial point (not at the point itself), and the transience means that
\[
\int_0^{\infty}\int_\Omega p_0(t,x,y)dydt<\infty,
\]
where $\Omega$ is a neighborhood of the point $x$. This condition implies that
\begin{equation}\label{rec1}
\int_1^{\infty}p_0(t,x,x)dt<\infty,
\end{equation}
Conditions (\ref{rec}),(\ref{rec1}), obviously, do not depend on $x$. The integrals (\ref{rec}),(\ref{rec1}) diverge for recurrent processes. The CLR estimates are valid for more general operators than (\ref{sc}) (see \cite{2,1,mv} and references there), but usually the transience is an essential requirement when these more general operators are considered.

Recall one of the forms (not the most general) of the CLR estimate for Schr\"{o}dinger-type  operators $H=H_0-V(x)$ on $L^{2}(X,%
\mathcal{B,}\mu )$ where
 $X$ is a complete $\sigma $-compact metric space with
Borel $\sigma $-algebra $\mathcal{B}(X)$ and a $\sigma $-finite measure $\mu
(dx).$ Let $H_{0}$ be a self-adjoint non-negative operator  such that the operator $-H_0$ is the generator of a Markov semigroup $P_t $ acting
on $C(X)$. Let $p_0(t,x,y)$ be the kernel of $P_t $, i.e., $p_0$ is the transition density of the underlying Markov process $x(t).$ If the process $x(t)$ is transient, then
\begin{equation} \label{1xx}
N_0(V) \leq \frac{1}{c(\sigma )}\int_{X}V(x)\int_{\frac{\sigma }{V(x)}%
}^{\infty }p_0(t,x,x)dt\mu (dx),~V\geq 0,
\end{equation}
where $\sigma >0$ is arbitrary and $c(\sigma )=e^{-\sigma }\int_{0}^{\infty }\frac{ze^{-z}dz}{z+\sigma }.$ Another widespread form of this estimate
\begin{equation}\label{c}
N_0(V)\leq C_{d}\int_{R^{d}}V^{\frac{d}{2}}(x)dx,~V\geq 0,
\end{equation}
for the Schr\"{o}dinger operator in $R^d,~d\geq 3$, follows immediately from (\ref{1xx}) since $p_0(t,x,x)=c_d t^{-d/2}$ in this case. This fact stresses the importance of the transiency requirement. It also explains the reason for imposing the assumption $d\geq 3$ in papers on CLR estimates.

This paper concerns the CLR-type estimates for operator (\ref{sc}) in dimensions $d=1$ and $2$ where the Markov process is recurrent and estimates (\ref{1xx}),(\ref{c}) fail. We also will consider the fractional-dimensional case $d\leq 2$ presented by the Bessel operators and the fractional degrees of the one-dimensional discrete Laplacian. As usual, the Laplacian on the lattice $Z^d$ is defined as follows
\begin{equation}\label{lapl}
\Delta \psi(x)=\sum_{x':|x-x'|=1}(\psi(x')-\psi(x)).
\end{equation}

The literature on the negative spectrum of Schr\"{o}dinger operators is so extensive that we are going to mention here only some of the most closely related papers. The quazi-classical asymptotics for $N_0(\alpha V)$,
$$
N_0(\alpha V)\sim c(d,V)\alpha^{d/2}, ~d\geq 1, ~~\alpha\to\infty,
$$
is valid \cite{rs} in dimensions $d\geq 3$ when $V\in L^{d/2}(R^d)$, and the CLR estimate (\ref{c}) is in agreement with the quazi-classical asymptotics. One could expect that $N_0(\alpha V)\sim c(V)\alpha,~\alpha\to\infty,$ when $d=2$ and $V\in L^1(R^2)$. However, while this is true \cite{rs} for ``good" potentials $V$, this asymptotics is not valid for arbitrary $V\in L^1(R^2)$ \cite {birS},\cite {birL},\cite{1}. One of our goals was to establish an estimate for $N_0(\alpha V)$ of the first or ``almost" first order in $\alpha$ when $d=2$. Thus we are not going to consider the Birman-Schwinger estimates \cite{bir}, \cite{sch}, \cite{rs} which are not sharp when $\alpha \gg 1$. The latter estimates, known for $d\geq 3$, were extended recently for low dimensions \cite{st}, but they have order $\alpha ^2$ when $d=2$. We will mostly focus on the two-dimensional case since one-dimensional problems are very specific and can be studied by a variety of methods. However, our general approach allows us also to obtain new one-dimensional results (for example, an improved Bargmann estimate and the Lieb-Thirring-type estimate with  $\gamma< 1/2$).

Our two-dimensional estimates for $N_0(\alpha V)$ have order $\alpha, \alpha\to \infty,$ in the lattice case and order $\alpha \ln \alpha$ for operators in $L^2(R^2)$. Simpler estimates of order $\alpha \ln \alpha$ for $N_0(\alpha V)$ were obtained in \cite{seto}, \cite{newton}, \cite{many} for two-dimensional operators with a central potential $V(x)=V(|x|)$. Using  a fundamental one-dimensional estimate of the Leib-Thirring sum from \cite{dirk} a remarkably simple formula is obtained in \cite{many}:
$$
N_0( V)\leq 1+\frac{1}{2\pi}\int_{R^2}V(|x|)|\ln|x||dx+\frac{1}{\pi\sqrt 3}\int_{R^2}V(|x|)dx.
$$
Unfortunately, this result is valid only for central potentials (a counter-example can be easily constructed if $V$ is not central). The authors also formulated an elegant conjecture for general $V$. The conjectured estimate contains a term with a central decreasing rearrangement of the potential. An estimate of order $\alpha$ in two dimensions for general potentials was obtained in \cite{solO} in terms of the local Orlicz norms of the potential (an earlier exposition of the author's technique and its development can be found in \cite{birS1}, \cite{birS2}). Our estimate in the continuous case is logarithmically weaker when $\alpha\to\infty$, but has a much simpler form.

Our approach to treat the recurrent operators is particularly simple when the dimension $d$ is less than $2$ and for the lattice two-dimensional operator. In these cases we replace the unperturbed operator $H_0=-\Delta$ by its rank one perturbation $H_1$ for which the killing (annihilation) of the corresponding Markov process in a single point is introduced. The idea to use the rank one perturbation in this problem is going to B. Simon \cite{sim11} and was used in \cite{st}. The difference is that we combine it with the CLR rather than with the Birman-Schwinger  estimates, and this allows us to get the result in several lines. After the killing is imposed, the Markov process generated by $H_1$ is transient. The standard CLR arguments can be applied to $H_1-V$, and therefore
\begin{equation} \label{1xxa}
N_0(V) \leq\frac{1}{c(\sigma )} \int_{X}V(x)\int_{\frac{\sigma }{V(x)}
}^{\infty }p_1(t,x,x)dt\mu (dx)+1 ,
\end{equation}
where $p_1$ is the solution of the problem
$$
\frac{dp_1}{dt}=-H_1 p_1,~ t>0,~~p_1(0,x,y)=\delta_y(x).
$$

Let us note that $p_0$ and $p_1$ are integrable near $t=0$ if $d<2$ and in the lattice case. Thus one can use the estimate (\ref{1xxa}) with $\sigma=0$ (and $c(0)=1$). This implies the following particular version of (\ref{1xxa}):
\begin{equation} \label{1xxab}
N_0(V) \leq\int_{X}V(x)\int_{0
}^{\infty }p_1(t,x,x)dt\mu (dx)+1.
\end{equation}
Consider, for example, a lattice Schr\"{o}dinger operator in dimensions one or two. Then $H_1$ is obtained by imposing the Dirichlet boundary condition at one point, for example, at $x=0$.
We will show that
\begin{eqnarray}
\sum_{x\in Z}p_1(t,x,x)&=&|x|,\quad d=1,\label{tran}\\
\sum_{x\in Z^2}p_1(t,x,x)&\leq & C\ln(2+|x|), \quad d=2. \label{tran1}
\end{eqnarray}
Thus, (\ref{1xxab}) implies
\begin{eqnarray}
N_0(V)&\leq&\sum_Z |x| V(x)+1,\quad d=1,\label{tranz}\\
N_0(V)&\leq & C\sum_{Z^2} \ln(2+|x|) V(x)+1, \quad d=2. \label{tran1z}
\end{eqnarray}

The continuous analog of the estimate (\ref{tranz}) (for the operator in $L^2(R)$) has the form
\begin{equation} \label{barg}
N_0(V)\leq\int_R |x| V(x)dx+1,
\end{equation}
and coincides with the well-known Bargmann estimate (see \cite{rs}).
The Bargmann estimate has a wrong scaling order ($\alpha$ instead of $\sqrt\alpha$). The estimates of order $\sqrt\alpha$ can be found in \cite{many}, \cite{nsol}, see also the Calogero estimate \cite{rs} for monotone potentials.

Inequality (\ref{1xxa}) with $\sigma >0$ leads to the following refined Bargmann's estimate for operator (\ref{sc}) in $L^2(R)$
\begin{equation} \label{rebarg}
N_0(V)\leq \frac{1}{c(\sigma )}[\int_{x^2V(x)>\sigma} |x| V(x)dx+\frac{1}{\sqrt{\sigma\pi}}\int_{x^2V(x)<\sigma} x^2 V^{3/2}(x)dx]+1,\quad d=1,
\end{equation}
with the some $c(\sigma)$ as in (\ref{1xxa}). Note that the Bargmann's estimate (\ref{barg}) as well as other mentioned above one-dimensional estimates do not provide any information in the case of the potential
\[
V(x)=O(\frac{1}{x^2 \ln|x|}), \quad |x|\to\infty,\quad d=1,
\]
(the integral in (\ref{barg}) diverges), while the refined formula (\ref{rebarg}) shows that $N_0(V)<\infty$ for this type of potentials.

The most technically difficult part of the paper concerns the Schr\"{o}dinger operator in $R^2$:
$$
H=-\Delta -V(x), \quad H_0=-\Delta, \quad V\geq 0, ~~~x\in R^2.
$$
A rank one perturbation approach does not work here. We replace $H_0$ by
$$
 H_1=-\Delta +q(x), \quad q=1~~ \text{for}~ |x|<1,~~q=0 ~~\text{for}~ |x|>1.
$$
The operator $H=-\Delta +q(x)-(V(x)+q(x))$ can be considered as the perturbation of $H_1$ by the potential $V+q$. The Birman-Schwinger principle implies that
$$
N_0(V)=N_0(V+q;H_1)\leq N_0(2V;H_1)+N_0(2q;H_1).
$$
where the second argument of the function $N_0$ is the unperturbed operator. We chose $q$ to be so small ($q\leq 1$) that $N_0(2q;H_1)\leq 1$, and therefore
$$
N_0(V)\leq N_0(2V;H_1)+1.
$$
We show that operator $H_1$ is transient, and moreover the following non-trivial estimate holds:
$$
p_1(t,x,x)\leq \frac {C(a)\ln^2(2+|x|)}{t\ln^2t}   \quad \text{when}~~ t>\gamma(x)=\max(1, a|x|^2\ln|x|), ~a>0.
$$
This implies that
$$
N_0(V)\leq \frac{1}{c(\sigma)}\int_{R^2} V(x)\int_{\frac{\sigma}{V(x)}}^\infty p_1(t,x,x)dt dx+1
$$
\begin{equation} \label{rebarg2}
\leq C_1(\sigma)\int_{V(x)<\frac{\sigma}{\gamma(x)}} \frac{V(x)}{\ln\frac{\sigma}{V(x)}}\ln^2(2+|x|)dx
+C_2(\sigma)\int_{V(x)>\frac{\sigma}{\gamma(x)}}V\ln\frac{\gamma(x)V}{\sigma}dx+1.
\end{equation}
A similar refined Bargmann-type estimate will be proved in the two-dimensional lattice case.

Let us stress that all the above estimates are not translation-invariant unlike the case of Schr\"{o}dinger operators in dimensions $d\geq 3$. The following arguments show that the estimates of the form
\begin{equation} \label{inv}
N_0(V)\leq\sum_{Z^d} V^\gamma(x)dx+c,
\end{equation}
for operators on $Z^d$ and similar estimates in the continuous case can not be valid in dimensions $d=1,2$. Consider the operator $H=-\Delta -\varepsilon \delta(x), ~x\in Z^d,~d=1$ or $2$. This operator with an arbitrary small $\varepsilon>0$ has exactly one negative eigenvalue \cite{many}. One can take an arbitrary sequence $\varepsilon=\varepsilon _n,~n=1,2... ,$ for which $\sum_n\varepsilon_n^\gamma\leq 1$ and choose a sequence $x=x_n$ to be so sparse that the eigenfunctions of the operators $H_n=-\Delta -\varepsilon_n \delta_{x_n}(x)$ are practically orthogonal, and the operator $H=-\Delta -\sum_n\varepsilon_n \delta_{x_n}(x)$ has infinitely many negative eigenvalues \cite{many}. Then the left-hand side of (\ref{inv}) is infinity and the right-hand side does not exceed $1+c$. Similar arguments will be provided for Schr\"{o}dinger operators on general lattices with the recurrent Markov process generated by the unperturbed operator (see section 6).

We will also obtain the estimates on $N_0(V)$ from below. The following results show that estimates (\ref{tranz})-(\ref{barg}) are sharp (relatively to the decay of the potential at infinity). Consider the lattice Schr\"{o}dinger operator on $Z^d$. Then $N_0(V)=\infty$ for two-dimensional operators with any potential $V$ such that
$$
\sum_{Z^2}V(x)=\infty
$$
and for any one-dimensional operator when
$$
\sum_{Z}\frac{ |x|}{\ln^{1+\varepsilon}(1+|x|)}V(x)=\infty \quad \text{for some}~~ \varepsilon>0.
$$
Similar results are valid in the continuous case. Note that a much stronger result is known \cite{GNY} for the continuous two-dimensional Schr\"{o}dinger operator:
$$
N_0(V)>C\int_{R^2}V(x)dx.
$$

Let us turn to non-integer dimensions. We will study two types of operators with fractional spectral dimension: Bessel operators and fractional powers of the lattice Laplacian. These are the operators on the half line $R_+$ and $R$, respectively, but their negative spectra behave as for multi-dimensional operators of dimension $d$ that is not necessarily an integer. The Bessel operators are defined by
\begin{equation} \label{bess}
B_d=\frac{d^2}{dr^2}+\frac{d-1}{r}\frac{d}{dr}=\frac{1}{r^{d-1}}\frac{d-1}{r}(r^{d-1}\frac{d}{dr}), \quad d\geq 1 \quad \text {can be not-integer,}
\end{equation}
in $L^2([0,\infty),~ r^{d-1}dr)$. Operator $B_d$ is selfadjoint when $d\geq 2$. If $d$ is strictly greater than $2$, then the diffusion process $b_d(t), t\geq 0,$ generated by $B_d$, is transient and the classical CLR estimates imply
$$
N_0(V)\leq C(d)\int_{0}^\infty  V^{d/2}(r)r^{d-1}dr.
$$
If $d=2$, then the process $b_d(t)$ is recurrent, and the estimates for $N_0(V)$ can be derived from the results obtained in this paper for two-dimensional Schr\"{o}dinger operators.

The process $b_d(t)$ as well as the operator $B_d$ are not determined by (\ref{bess}) if $d<2$. One needs to add a boundary condition at $r=0$. The operator $B_d$ and the process $b_d(t)$ become well-defined if the Dirichlet boundary condition is imposed at  $r=0$, and the following analog of (\ref{rebarg}) is valid in this case:
$$
N_0(V)\leq c_1(\sigma)\int_{r:r^2V>\sigma}V(r)r^{2-d}dr+c_2(\sigma)\int_{r:r^2V<\sigma}V^{2-d/2}(r)r^{4-2d}dr.
$$

Somewhat similar results are valid for the fractional powers of the lattice operator. We will only consider the powers $(-\Delta)^\alpha$ of the one-dimensional lattice Laplacian. It will be shown that the Bargmann-type estimates are valid in this case with the constant $d=1/\alpha$ playing the role of the spectral dimension. Note that the Markov process generated by the operator $-(-\Delta)^\alpha$ is not local (it has a positive probability of jumping to any point of the lattice). The results for the operator$(-\Delta)^\alpha$ will be obtained as a consequence of the results for more general  Schr\"{o}dinger operators on discrete graphs which will be studied in section 6.

The final section of the paper is devoted to estimates on the Lieb-Thirring sums
$$
S_\gamma(V)=\sum_{i:\lambda_i\leq 0}|\lambda_i^\gamma|.
$$
The classical result \cite {Lt1} for the Schr\"{o}dinger operators in $R^d$ has the form:
\begin{equation} \label{lith}
S_\gamma(V)\leq c_{d,\gamma}\int_{R^d}V^{\frac{d}{2}+\gamma}dx,~~\frac{d}{2}+\gamma>1.
\end{equation}
This formula does not cover two cases: $d=2, \gamma=0$ and $d=1, \gamma< 1/2$. The estimate (\ref{lith}) in the borderline case $d=1, \gamma=1/2$ with a sharp constant was obtained in \cite{dirk}. Similar estimate for Jacobi matrices, which include the lattice one-dimensional Laplacian as a particular case, was proved in \cite{hs}.

Our approach based on the annihilation of the underlying recurrent process allows us to obtain estimates on $S_\gamma(V)$ for $d=1,2$.
Here we provide only two simple results. Other statements and a discussion are in section 9. The advantage of our estimate in the case  $d=2, \gamma\in[0,1]$ is related not to the fact that $\gamma$ can be zero, but to the independence of the constant from $\gamma$, contrary to (\ref{lith}) where $c_{d,\gamma}\to\infty$ as $\gamma\to 0.$

Consider one-dimensional operator (\ref{sc}) with a bounded potential: $V(x)\leq \Lambda<\infty$. Then for $\gamma<1/2$ we have
$$
S_\gamma(V)\leq \Lambda^\gamma +\beta(\gamma)\int_{-\infty}^\infty V(x)|x|^{1-2\gamma}dx,~~~\beta(\gamma)=\pi^{-1/2}\gamma \Gamma(\gamma)\int_0^\infty\frac{1-e^{\frac{-1}{s}}}{s^{1/2+\gamma}}ds.
$$
Here $\Gamma(\gamma)$
is the gamma-function. This is a Bargman-type estimate for $
S_\gamma(V)$. We will provide also a refined Bargman-type estimate. The next estimate is not very sharp in some cases (in particular it is much worse than in \cite{dirk} when $\gamma=1/2$), but it is valid for each $\gamma>0$:
\begin{equation} \label{rebarg11}
S_\gamma(V)\leq \Lambda^\gamma +\int_{-\infty}^\infty |x| V^{1+\gamma}(x)dx,\quad d=1.
\end{equation}

The following estimate will be obtained for the two-dimensional Schr\"{o}dinger operators: if $0\leq V(x)\leq 1$, then there are some constants $a_1,a_2$ such that for each $\gamma\in[0,1]$,
$$
S_\gamma(V)\leq a_1+a_2\int_{R^2}\frac{V^{1+\gamma}(x)}{\ln\frac{4}{V(x)}}\ln^2(2+|x|)dx.
$$
Its analog is also valid in the lattice case without the assumption of the boundedness of the potential.

The paper is organized as follows. Section 2 is devoted to the one-dimensional discrete and continuous Schr\"{o}dinger operators. Two-dimensional continuous and lattice  Schr\"{o}dinger operators will be studied in sections 3 and 4, respectively. Proofs of some important lemmas will be given in the Appendix. Estimates from below are given in section 5 showing the sharpness of the results in sections 2-4. Operators on general discrete graphs are considered in section 6. In particular, it is shown there that one can't expect the translation-invariant estimates for $N_0(V)$ to be valid when the underlying Markov process is recurrent. Sections 7 and 8 concern the fractional powers of the lattice Laplacian and Bessel operators. Lieb-Thirring estimates are studied in section 9.

The authors are grateful to O. Safronov for productive discussions and to B. Simon for useful critical remarks.

\section{One-dimensional operators}

\begin{theorem} The Bargmann (\ref{barg}) and refined Bargmann (\ref{rebarg}) estimates hold for the operator
$$
H=-\frac{d^2}{dx^2}-V(x) \quad \text {in}~~ L^2(R).
$$
\end{theorem}
\textbf{Remark.} It was mentioned in the introduction that estimate (\ref{barg}) is well-known.

 \textbf{Proof.} Consider a rank one perturbation $H_1$ of the operator $H_0=-\frac{d^2}{dx^2}$ which is obtained by imposing the Dirtichlet boundary condition at $x=0:$
$$
H_1=-\frac{d^2}{dx^2}, \quad  D_{H_1}=\overline{C_0^\infty(R)\bigcap \{y:y(0)=0\}}.
$$
Then
$$
N_0(V;H_0)\leq N_0(V;H_1)+1\leq \frac{1}{c(\sigma)}\int_R V(x)\int_{\frac{\sigma}{V(x)}}p_1(t,x,x)dt dx+1,
$$
where
$$
\frac{d}{dt}p_{1}=\frac{d^2}{dx^2}p_1, ~~~t>0;~~~p_1(t,0,y)=0,~~~p_1(0,x,y)=\delta_y(x).
$$
Then
$$
p_1(t,x,y)=\frac{e^{-\frac{(x-y)^2}{4t}}}{\sqrt{4\pi t}}-\frac{e^{-\frac{(x+y)^2}{4t}}}{\sqrt{4\pi t}},~~x,y>0.
$$
Thus
\begin{equation} \label{p11a}
p_1(t,x,x)=\frac{1-e^{-\frac{x^2}{4t}}}{\sqrt{4\pi t}}.
\end{equation}
Similarly, the kernel $R^{(1)}_\lambda (x,y)$ of the resolvent $(-H_1-\lambda)^{-1}$ satisfies
$$
R^{(1)}_\lambda (x,y)=\frac{e^{-\sqrt\lambda|x+y|}-e^{-\sqrt\lambda|x-y|}}{2\sqrt\lambda}, \quad R^{(1)}_{+0} (x,x)=-|x|.
$$

Since $\int_0^\infty p_1(t,x,x)dt=-R^{(1)}_{+0}$, the latter relation together with (\ref{1xxab}) imply (\ref{barg}). Inequality (\ref{rebarg}) immediately follows from (\ref{1xxa}) and (\ref{p11a}) since the formula for $p_1$ above implies that
$$
\int_0^\infty p_1(t,x,x)dt=|x|F(\frac{\sigma}{V(x)x^2}), \quad F(\gamma)=\int_\gamma^\infty \frac{1-e^{\frac{-1}{4\tau}}}{\sqrt{4\pi\tau}}d\tau,
$$
and $F(\gamma)\leq 1$ for all $\tau\geq 0;$ $F(\gamma)\leq \int_\gamma^\infty \frac{{1}}{4\tau\sqrt{4\pi\tau}}d\tau=\frac{1}{4\sqrt{\pi\gamma}} $ when $\gamma\geq 1.$
\qed

Consider now the same operator on the one-dimensional lattice:
$$
H\psi(x)=-\Delta \psi-V(x) \psi=2\psi(x)-\psi(x+1)-\psi(x-1)-V(x)\psi(x) \quad \text {in}~~ L^2(Z).
$$
 The general solution of the equation $\Delta \psi-\lambda \psi =0,~\lambda >0,$ on the lattice $Z$ has the form $\psi =C_1a_1^x+C_2a_2^x$, where $a_{1,2}$ are the roots of the equation $a^2-(2+\lambda )a +1=0.$ If $a=\frac{2+\lambda+\sqrt{\lambda ^2+4\lambda}}{2},~\lambda >0,$ is the biggest root, then the solution of the equation
$$
(\Delta -\lambda ) R_{\lambda}^{(0)}(x,y) =\delta (x-y)
$$
must have the form
$R_{\lambda}^{(0)}(x,y)=ca^{-|x-y|}$ where the constant $c$ can be easily found from the equation.  This leads to
\[
R_{\lambda}^{(0)}(x,y) =\frac{a^{1-|x-y|}}{2-(2+\lambda)a},\quad \lambda >0.
\]
If $H_1$ is the lattice Laplacian with the Dirichlet boundary condition at $x=0$ and
$$
R_{\lambda}^{(1)}(x,y)=-\int_0^\infty p_1(t,x,y)e^{-\lambda t}dt
$$
is the kernel of its resolvent, then $R_{\lambda}^{(1)}(x,y)=R_{\lambda}^{(0)}(x,y)-R_{\lambda}^{(0)}(x,-y)$, and
\[
R_{\lambda}^{(1)}(x,x)=\frac{a-a^{1-2|x|}}{2-(2+\lambda)a},\quad \lambda >0.
\]
We note that $a \sim 1+ \sqrt{\lambda}$ and $2-(2+\lambda)a \sim -2\sqrt{\lambda}$ as $\lambda \to +0$. Hence,
$-R_{0}^{(1)}(x,x)=|x|$, and  therefore (\ref{tran}), (\ref{tranz}) are proved for the one-dimensional lattice operator.

In order to obtain a refined Bargmann estimate in the lattice case, we note that
$$
p_1(t,x,y)=p_0(t,x,y)-p_0(t,x,-y), \quad \text{where}~~~p_0(t,x,y)=\frac{1}{2\pi}\int_{-\pi}^\pi e^{-2t(1-\cos\phi)+i(x-y)\phi}d\phi,
$$
i.e.,
$$
p_1(t,x,x)=p_0(t,x,x)-p_0(t,x,-x)=p_0(t,0,0)-p_0(t,2x,0).
$$
The integral above can be expressed through the modified Bessel function. This allows to obtain the asymptotic behavior of $p_0(t,x,0)$ as $t,|x|\to\infty$. Another option is to apply Cramer's form of the central limiting theorem \cite{F} (Ch. 16, 7) which leads to the following result: if $t\to\infty$ then
$$
p_0(t,x,0)= \frac{ e^{-\frac{x^2}{4t}+O(\frac{|x|^4}{t^3})}}{\sqrt{4\pi t}}(1+O(\frac{1}{t})), \quad \text{for} \quad |x|\leq t^{2/3},
$$
$$
|p_0|\leq e^{-ct^{1/3}},~~~|x|\geq t^{2/3}.
$$
These formulas allow us to obtain the same estimate for $\int_\gamma^\infty p_1(t,x,x)dx$ as in the continuous case, which leads to
$$
N_0(V)\leq C_1(\sigma)\sum_{x:V(x)>\frac{\sigma}{x^2}}|x|V(x)+C_2(\sigma)\sum_{x:V(x)<\frac{\sigma}{x^2}}x^2V^{\frac{3}{2}}(x)+1.
$$

\section{Two-dimensional Schr\"{o}dinger operator, continuous case}

This section is devoted to the estimate of $N_0(V)$ for the two-dimensional operator (\ref{sc}) in
$L^2(R^2).$
The rank one perturbation does not work in this case. We consider a soft killing by a potential instead. Let  $H_1=-\Delta+q(x),~x\in R^2,$ be the perturbation of the operator $H_0=-\Delta$, where $q(x)=1$ when $|x|<1,~q(x)=0$ when $|x|\geq 1.$
Let $p_1=p_1(t,x,y)$ be the solution of the corresponding parabolic problem
\begin{equation} \label{p12}
p_{1_t}=\Delta p_1-q(x)p_1,~~t>0,~~p_1(0,x,y)=\delta_y(x).
\end{equation}
We need to show that the Markov process with the generator $H_1$ is transient and we need a sharp estimate of $p_1$ as $t\to\infty$.
\begin{theorem}\label{thp1}
The following estimate holds
\begin{equation} \label{p1loc}
|p_{1}(t,x,x)|\leq\frac{C(a)\ln^2|x|}{t\ln^2 t}~~~\text{when}~~t>\gamma(x)=\max(1,a|x|^2\ln|x|)
\end{equation}
for some $a>0.$
\end{theorem}
\textbf{Proof.} We will provide here only a sketch of the proof. The rigorous arguments will be given in the Appendix, Lemmas \ref{ogr}-\ref{ogr3}. The first of these lemmas states that
\begin{equation} \label{ap1loc}
|p_{1}(t,x,y)|\leq\frac{C}{t\ln^2 t},~~|x|,|y|\leq 2,~~t>2.
\end{equation}

In order to justify this estimate we solve (\ref{p12}) using the Laplace transform and arrive at
\begin{equation} \label{aGa}
p_1=-\int_{\Gamma} R_{\lambda}^{(1)}(x,y)e^{\lambda t}d\lambda, \quad x \neq y,
\end{equation}
where $R^{(1)}_\lambda(x,y)$ is the kernel of the resolvent
$$
R^{(1)}_\lambda=(\Delta-q(x)-\lambda)^{-1},
$$
and the contour $\Gamma$ consists of the bisectors of the third and second quadrants of the $\lambda-$plane with the  direction on $\Gamma$ such that Im$ \lambda$ increases when a point moves along $\Gamma$. We show (in the proof of Lemma \ref{ogr}) that the kernel
 $R^{(1)}_\lambda(x,y)$ is bounded at $\lambda=0$, and
\begin{equation} \label{alto0}
R^{(1)}_\lambda(x,y)=a(x,y)+\frac{1}{\ln\lambda}b(x,y)+O(\frac{1}{\ln^{2}\lambda}), \quad \lambda\to 0.
\end{equation}

Obviously,
\begin{equation} \label{t1t}
\int_\Gamma e^{\lambda t}d\lambda=0 \quad \text{for}~~ t>0.
\end{equation}
Further, replacing $\Gamma$ be a contour $\gamma$ around the negative semi-axis in the $\lambda-$plane we obtain that
\[
\int_{\Gamma}\frac{1}{\ln\lambda} e^{\lambda t}d\lambda=\int_{\gamma}\frac{1}{\ln\lambda} e^{\lambda t}d\lambda
=\int_0^{\infty}[\frac{1}{\ln\sigma+\pi i}-\frac{1}{\ln\sigma-\pi i}]e^{-\sigma t}d\sigma
\]
\begin{equation} \label{tt}
=\int_0^{\infty}\frac{2\pi i}{\ln^2\sigma+\pi^2}e^{-\sigma t}d\sigma=\frac{2\pi i}{t\ln^2t}+O(\frac{1}{t\ln^3t})
,~~t>2.
\end{equation}
The last two relations together with (\ref{alto0}) and (\ref{aGa}) imply (\ref{ap1loc}).

Lemma \ref{ogr2} extends (\ref{ap1loc}) for arbitrary $x$ if $t$ is large enough. It states that
\begin{equation} \label{ap1loc21}
|p_{1}(t,x,y)|\leq\frac{C(a)\ln(2+|x|)}{t\ln^2 t},\quad ~~|y|\leq 3/2, |x|\geq 2 \quad \text{if} ~~t\geq \max (1,a|x|^{2}\ln|x|)
\end{equation}
for some $a>0.$ Note that $\ln(2+|x|)$ appears in the estimate for $p_1$, and the estimate is valid only if $t\gg |x|^2$.
The main step in the proof is a comparison of $p_1$ with the function $A\psi(t,x)+b\phi(t,x)$,  where
\begin{equation} \label{a1a}
\psi(t,x)=\int_\Gamma v(\lambda,x)e^{\lambda t}d\lambda, \quad v=v(\lambda,x)=\frac{K(\sqrt\lambda |x|)}{\ln \lambda}.
\end{equation}
Here
\[K(\mu)=K_0(\mu)=\frac {\pi i}{2}H^{(1)}_0(i\mu),
~ \mu>0,
\]
is the modified Bessel function (it is proportional to the Hankel function of the pure imaginary argument).

Function $\psi$ satisfies the heat equation when $|x|\geq 2$, vanishes when $t=0, |x|\geq 2$, and
\begin{equation} \label{apsi1}
\psi|_{|x|=2}=\frac{c_1}{t\ln^2 t}+O(\frac{1}{t\ln^3 t}), \quad t\to \infty,
\end{equation}
since expansion (\ref{alto0}) holds for $v$ when $|x|=2$.
Thus there exist constants $A$ and $\tau$ such that $A\psi > p_1$ when $|x|=2, t\geq \tau.$ We choose $\phi$ to be the solution of the heat equation for $|x|>2$ with zero initial data and the boundary condition at $|x|=2$ being zero for $t>\tau$ and one for $t<\tau$. If $b$ is large enough, then $p_1<A\psi(t,x)+b\phi(t,x)$ at $|x|=2,$ and therefore $p_1<A\psi(t,x)+b\phi(t,x)$ at $|x|>2.$ It remains to estimate functions $\psi$ and $|\phi|$ for $|x|\gg2$.

The asymptotic behavior of $\psi$ for large values of $t$ depends on the behavior of the integrand in (\ref{a1a}) as $\lambda\to 0$. If we use only the main terms of the expansion of the modified Bessel function at zero, we obtain that
$$
\psi\sim \int _\Gamma \frac{a+b\ln(\sqrt\lambda |x|)}{\ln\lambda}e^{\lambda t}d\lambda, \quad t\to\infty.
$$
After that, (\ref{t1t}) and (\ref{tt}) imply (\ref{ap1loc21}). A similar estimate is valid for $\phi$.

The last lemma in the Appendix extends (\ref{ap1loc21}) (with an extra logarithmic factor in the right-hand side of the inequality) to arbitrary $x$ and $y$ and completes the proof of the theorem.
\qed

This sketch of the proof avoids some difficulties and it does not explain the reason to introduce the condition $t\gg |x|^2$. The rigorous proof will be given in the Appendix. A similar result in the lattice case will be proved in the next section using probabilistic ideas.
\begin{theorem} \label{tn2}
Estimate (\ref{rebarg2}) holds for two-dimensional Schr\"{o}dinger operators (\ref{sc}) in $L^2(R^2)$.
\end{theorem}
\textbf{Proof.} The Birman-Schwinger principle implies that
$$
N_0(V)=N_0(V+q;H_1)\leq N_0(2V;H_1)+N_0(2q;H_1).
$$

Let us show that $N_0(2q;H_1)\leq 1.$ Indeed, operator $H_1$ perturbed by the potential $-2q$ coincides with the operator $-\Delta-q(x)$, i.e., one needs to show that the latter operator has at most one negative eigenvalue. In fact it has exactly one eigenvalue, but we need only the estimate from above. The eigenfunctions $\psi_n(x)$ of this operator can be found by separation of variables, i.e., they have the form
$$
\psi_n=(\alpha_n \cos(n\theta)+\beta_n \sin(n\theta))f_n(r), ~~n\geq 0,
$$
where $r=|x|,~\theta=\arctan(y/x)$ and $(-\Delta-q(x))f_n=\gamma_nf_n$. The corresponding eigenvalues are $\lambda_n=n^2+\gamma_n.$ Since the operator $-\Delta$ is strictly positive and $q(x)\leq 1$, we have $\gamma_n>-1$. Thus $\lambda_n>0$ when $n>0$ and $\lambda_0>-1$. In fact, the operator $-\Delta-q(x)$ can not have positive eigenvalues, i.e., this operator has only spherically symmetrical eigenfunctions $f=f_0(r)$ and the corresponding eigenvalues $\lambda \in(-1,0]$.

Let us show that the eigenvalue problem $(-\Delta-q(x))f_0=\lambda f_0$ has at most one eigenvalue with a spherically symmetrical eigenfunction. We write the problem in polar coordinates:
\begin{equation} \label{onepr}
-y''(r)-\frac{1}{r}y'(r)-q(r)y(r)=\lambda y(r), ~~~\lambda\in(-1, 0],~~~y(0)<\infty, ~~\int_0^\infty ry^2dr<\infty.
\end{equation}
The equation does not have nonzero solutions in $L^2(R_+)$ if $\lambda=0$. If $y$ satisfies (\ref{onepr}) with $-1<\lambda<0$, then
$$
y=C_1J_0(\sqrt{1+\lambda}r),~~r\leq 1;~~~y=C_2K_0(\sqrt{|\lambda}|r),~~r\geq 1,
$$
where $J_0$ is the Bessel function and $K_0$ is the modified Bessel function (Hankel function of the purely imaginary argument). Since $K_0(\sigma)\neq 0$ when $\sigma>0$ and $J_0(\sigma)\neq 0$ when $1>\sigma>0$, the function $y$ does not vanish at $r>0.$ If there exist two nonzero solutions of problem (\ref{onepr}) with different values of $\lambda$, then at least one of them must change its sign on $R_+$. Hence, $N_0(2q;H_1)\leq 1,$ and the first inequality in (\ref{rebarg2})
is proved.

In order to prove the second inequality in (\ref{rebarg2}) we rewrite the first inequality in the form
$$
N_0(V)\leq c(\sigma)\int_{D_1} V(x)\int_{\frac{\sigma}{V(x)}}^\infty p_1(t,x,x)dt dx+c(\sigma)\int_{D_2}\int_{\frac{\sigma} {V(x)}}^\infty  V(x)p_1(t,x,x)dt dx+1,
$$
where $D_1=\{x:\frac{\sigma}{V(x)}>\gamma (x)\},~D_2=R^2\setminus D_1$ and $\gamma(x)$ is defined in  (\ref{rebarg2}) or (\ref{p1loc}) with $a=1.$ Theorem \ref{thp1} allows us to rewrite the first term in the right-hand side of the inequality above as the first term in (\ref{rebarg2}). Hence, in order to prove the second inequality in (\ref{rebarg2}) it remains to show that
\begin{equation} \label{above}
\int_{\frac{\sigma}{V(x)}}^\infty p_1(t,x,x)dt \leq C\ln\frac{(2+|x|)V}{\sigma},~~~x\in D_2.
\end{equation}

We split the interval of integration in two parts $(\frac{\sigma}{V(x)},1+\gamma(x))\bigcup (1+\gamma(x),\infty)$. We estimate function $|p_1|$ from above on the first interval by $1/4\pi t$ and apply Theorem \ref{thp1} on the second interval. This implies that the left-hand side in (\ref{above}) does not exceed
$$
\ln(1+\gamma)-\ln\frac{\sigma}{V(x)}+\frac{C\ln^2(2+|x|)}{\ln(1+\gamma)}.
$$
The latter value can be estimated by the right-hand side  in (\ref{above}) since
$$
C_1\ln(1+\gamma)\leq \ln(2+|x|)\leq C_2\ln(1+\gamma).
$$

\qed

\section{A two-dimensional lattice operator}

We consider two-dimensional lattice operators
$$
H\psi(x)=-\Delta \psi-V(x) \psi \quad \text {in}~~ L^2(Z^2)
$$
in this section. First we will prove (\ref{tran1}) which justifies the Bargmann estimate (\ref{tran1z}) ($\sigma =0$). Then we will prove the refined estimate ($\sigma >0$).

The following two facts are the key starting points for the proof of (\ref{tran1}). Let $R_{\lambda}^{(0)}(x,y),$ $ R_{\lambda}^{(1)}(x,y)$ be the kernels of the resolvents of the operators $\Delta =-H_0$ and $-H_1$, respectively, where $-H_1$ is obtained from $-H_0$ by imposing the Dirichlet boundary condition at the origin (annihilation of the Markov process at this point). Obviously,
\[
R_{\lambda}^{(1)}(x,y)=-\int_0^{\infty}p_1(t,x,y)e^{-\lambda t}dt,~~\lambda >0,
\]
where $p_1$ is the transition probability for the Markov process with the generator $-H_1$.
Since $p_1\geq 0,$ one can pass to the limit as $\lambda \to 0$ in the relation above.
Thus, the transience of the process $x(t)$ for the operator $H_1$ is equivalent to the condition $|R_{0}^{(1)}(x,x)|<\infty$, and it is sufficient to prove estimate (\ref{tran1}) for $-R_{0}^{(1)}(x,x)$. Secondly, the following relation is valid for the latter function:
\begin{equation} \label{res}
R_{0}^{(1)}(x,x)=2\lim_{\lambda \to +0}[R_{\lambda}^{(0)}(0,0)-R_{\lambda}^{(0)}(x,0)].
\end{equation}
Indeed, the kernel $R_{\lambda}^1(x,y),~\lambda >0,$ must have the form  $R_{\lambda}^1(x,y)=R_{\lambda}^0(x,y)+cR_{\lambda}^0(x,0)$, where $c=c(y)$ can be found from the condition $R_{\lambda}^1(0,y)=0$. This immediately implies
\begin{equation} \label{res1ab}
R_{\lambda}^{(1)}(x,x)=[R_{\lambda}^{(0)}(x,x)-\frac{[R_{\lambda}^{(0)}(x,0)]^2}{R_{\lambda}^{(0)}(0,0)}],~~\lambda>0,
\end{equation}
and
\begin{equation} \label{res1}
R_{0}^{(1)}(x,x)=\lim_{\lambda \to +0}[R_{\lambda}^{(0)}(x,x)-\frac{[R_{\lambda}^{(0)}(x,0)]^2}{R_{\lambda}^{(0)}(0,0)}].
\end{equation}
Note that (\ref{res1}) holds for general discrete operators.

Formula (\ref{res1}) for general discrete operators can be written in the form (\ref{res}) when the following three conditions hold: operator $H_0$ is translation invariant, the Markov process with the generator $-H_0$ is recurrent, and the Markov process with the generator $-H_1$ is transient (in particular, if $H_0=-\Delta$ on $Z^2$). Indeed, (\ref{res1ab}) can be rewritten in the form
$$
R_{\lambda}^{(1)}(x,x)=\frac{B(R_{\lambda}^{(0)}(0,0)+R_{\lambda}^{(0)}(x,0))}
{R_{\lambda}^{(0)}(0,0)},
$$
where $B=R_{\lambda}^{(0)}(0,0)-R_{\lambda}^{(0)}(x,0)$. Since $R_{\lambda}^{(0)}(0,0)$ and $R_{\lambda}^{(0)}(x,0)$ have the same sign (they are negative), the ratio satisfies
$$\frac
{R_{\lambda}^{(0)}(0,0)}{(R_{\lambda}^{(0)}(0,0)+R_{\lambda}^{(0)}(x,0))}\in[0,1].
$$
Hence $B$ is bounded when $\lambda\to 0$. From here and $\lim_{\lambda \to +0}|R_{\lambda}^{(0)}(0,0)|=\infty, ~|R_{0}^{(1)}(x,x)|<\infty$ it follows that
$$
R_{0}^{(1)}(x,x)=\lim_{\lambda \to +0}\frac{[R_{\lambda}^{(0)}(0,0)]^2-[R_{\lambda}^{(0)}(x,0)]^2}{R_{\lambda}^{(0)}(0,0)}=\lim_{\lambda \to +0}\{B[2-\frac{B}{R_{\lambda}^{(0)}(0,0)}]\}=2\lim_{\lambda\to 0}B,
$$
i.e., (\ref{res}) holds.
\begin{theorem} \label{t1}
Let $H_1=-\Delta$ be the negative Laplacian in $L^2(Z^2)$ with the Dirichlet boundary condition at $x=0$. Then relations (\ref{tran1}), (\ref{tran1z}) hold.
\end{theorem}
\textbf{Proof.} We will show that for each fixed $x\in Z^2$,
\begin{equation} \label{res2}
R_{\lambda}^{(0)}(x,0)=\frac{1}{2\pi}\ln \lambda+u(x)+o(1) \quad \text{as} \quad \lambda \to +0, \quad \text{where}
 \quad |u(x)|\leq C\ln(2+|x|).
\end{equation}
This and (\ref{res}) imply that $
R_{0}^{(1)}(x,x)=2[u(0)-u(x)]$, and therefore
$|R_{0}^{(1)}(x,x)|\leq C\ln(2+|x|)$. The latter is equivalent to (\ref{tran1}) and justifies (\ref{tran1z}). Thus it remains only to prove (\ref{res2}).

The Fourier method applied to the equation $(\Delta-\lambda)\psi=\delta(x)$ leads, for $\lambda>0$, to
$$
R_{\lambda}^{(0)}(x,0)=\frac{1}{(2\pi)^2}\int_{[\pi,\pi]^2}\frac{e^{i(x,\phi)}d\phi}{2\cos\phi_1+2\cos\phi_2-4-\lambda}
$$
\begin{equation} \label{resla}
=
\frac{-1}{(2\pi)^2}\int_{[-\pi,\pi]^2}\frac{e^{i(x,\phi)}d\phi}{4\sin^2\frac{\phi_1}{2}+4\sin^2\frac{\phi_2}{2}+\lambda},
\end{equation}
where $\phi=(\phi_1,\phi_2)\in [-\pi,\pi]^2 \subset R^2.$ We put here
$$
4\sin^2\frac{\phi_1}{2}+4\sin^2\frac{\phi_2}{2}=|\phi|^2+h(\phi),~|h(\phi)|<C|\phi|^4.
$$
The difference between (\ref{resla}) and the same integral with $h(\phi)=0$ is
\[
\frac{1}{(2\pi)^2}\int_{[-\pi,\pi]^2}\frac{h(\phi)e^{i(x,\phi)}d\phi}
{[4\sin^2\frac{\phi_1}{2}+4\sin^2\frac{\phi_2}{2}+\lambda][|\phi|^2+\lambda]}.
\]
The latter integral converges to a bounded function
\[
v(x)=\frac{1}{(2\pi)^2}\int_{[-\pi,\pi]^2}\frac{h(\phi)e^{i(x,\phi)}d\phi}
{4[\sin^2\frac{\phi_1}{2}+\sin^2\frac{\phi_2}{2}]|\phi|^2}
\]
as $\lambda\to +0,$
 i.e.,
\[
R_{\lambda}^{(0)}(x,0)=\frac{-1}{(2\pi)^2}\int_{[-\pi,\pi]^2}\frac{e^{i(x,\phi)}d\phi}{|\phi|^2+\lambda}+v(x)+o(1), \quad
\lambda\to +0, \quad |v(x)|<C.
\]
Obviously, the function
\[
v_1(x)=\lim_{\lambda \to +0}\int_{[-\pi,\pi]^2\setminus \{|\phi|>1\}}\frac{e^{i(x,\phi)}d\phi}{|\phi|^2+\lambda}
\]
is also bounded in $x$. Hence,
\begin{equation} \label{11}
R_{\lambda}^{(0)}(x,0)=\frac{-1}{(2\pi)^2}\int_{|\phi|<1}\frac{e^{i(x,\phi)}d\phi}{|\phi|^2+\lambda}+w(x)+o(1), \quad
\lambda\to +0,  \quad |w(x)|<C.
\end{equation}

Note that
\[
\int_{|\phi|<1}\frac{d\phi}{|\phi|^2+\lambda}=\pi[\ln(1+\lambda)-\ln\lambda].
\]
Thus (\ref{11}) implies (\ref{res2}) if it is shown that
\begin{equation} \label{21}
|F(x)|<C\ln(2+|x|),
\end{equation}
where
\[
F=\lim_{\lambda \to +0}\int_{|\phi|<1}\frac{[e^{i(x,\phi)}-1]d\phi}{|\phi|^2+\lambda}
=\int_{|\phi|<1}\frac{[e^{i(x,\phi)}-1]d\phi}{|\phi|^2}
\]
The function $F$, considered for all $x\in R^2$, depends only on $r=|x|$. One can replace $x$ in the formula above by $x=(r,0)$. Then after passing to the polar coordinates $\sigma=|\phi|,~ \theta=\arctan \phi_2/\phi_1$, we obtain
\[
\frac{dF}{dr}=i\int_{|\phi|<1}\frac{\phi_1e^{ir\phi_1}d\phi}{|\phi|^2}=i\int_0^{2\pi}\int_0^1 \cos \theta e^{ir\sigma\cos \theta }d\sigma d\theta=\frac{2\pi}{r}.
\]

This justifies (\ref{21}) and completes the proof of the theorem.
\qed

The next statement provides the lattice analog of inequality (\ref{rebarg2}).
\begin{theorem}\label{t42}
The following estimate is valid for the number of negative eigenvalues of the lattice two-dimensional Schr\"{o}dinger operators:
\begin{equation} \label{rebarg2l}
N_0(V)\leq C_1(\sigma)\sum_{x:V(x)<\frac{\sigma}{\gamma(x)}} \frac{V(x)}{\ln\frac{\sigma}{V(x)}}\ln^2(2+|x|)
+C_2(\sigma)\sum_{x:V(x)>\frac{\sigma}{\gamma(x)}}V\ln\frac{\gamma(x) }{\sigma}+1,
\end{equation}
where $\gamma(x)=\max(1,|x|^2\ln|x|).$
\end{theorem}
\textbf{Remarks. 1.} The main difference between (\ref{rebarg2l}) and (\ref{rebarg2}) is that the second integrand in
 (\ref{rebarg2}) contains $V$ under the logarithm sign which is absent in (\ref{rebarg2l}). Its presence in  (\ref{rebarg2}) is due to the non-integrability of the transition probability $p_0(t,x,x)$ at $t=0$ for the Laplacian in $R^2$.

\textbf{ 2.}  After  (\ref{rebarg2l}) is proved, one can get a better estimate:
\begin{equation} \label{rebarg2la}
N_0(V)\leq C_1(\sigma)\sum_{x:V(x)<\frac{\sigma}{\gamma(x)}} \frac{V(x)}{\ln\frac{\sigma}{V(x)}}\ln^2(2+|x|)
+C_2(\sigma)\sum_{x:1>V(x)>\frac{\sigma}{\gamma(x)}}V\ln\frac{\gamma(x) }{\sigma}+N+1,
\end{equation}
where $N=\#\{x:V(x)\geq 1\}$. Indeed, let us introduce the potential $\widetilde{V}(x)$ which coincides with $V$ at the points $x$ where $V(x)<1$, and $\widetilde{V}(x)=0$ elsewhere. The operators $H$ with the potentials $V$ and $\widetilde{V}$ differ by an operator of rank $N$, and the difference between the numbers of their eigenvalues can be at most $N$. Thus estimate (\ref{rebarg2l}) for the potential $\widetilde{V}$ implies (\ref{rebarg2la}).

\textbf{Proof.} We only need to show that
\begin{equation} \label{ababa}
p_1(t,x,x)\leq \frac{C\ln^2|x|}{t\ln^2t} \quad \text{when} \quad t>\gamma(x),
\end{equation}
where $p_1(t,x,y)$ is the transition probability of the Markov process generated by $\Delta$ with the annihilation at the origin $x=0:~p_1(t,x,x)=P_x\{x(t)=x,x(s)\neq0,s\in[0,t]\}.$ The estimate (\ref{rebarg2l}) follows immediately from (\ref{ababa}) and (\ref{1xxa}). One can use the same approach to prove (\ref{ababa}) as in the case of operators in $R^2$, using the representation of $p_1$ through the inverse Laplace transform of the resolvent $R_{\lambda}^{(1)}(x,x)$ followed by the asymptotic analysis the corresponding integral. We decided to describe another approach to justify (\ref{ababa}) using probabilistic ideas. However, we will provide only a sketch of the proof.

We use formula (\ref{res1ab}) for the resolvent
$$
R_{\lambda}^{(1)}(x,x)=-\int_0^\infty e^{-\lambda t}p_1(t,x,x)dt.
$$
The ratio $\frac{R_{\lambda}^{(0)}(x,0)}{R_{\lambda}^{(0)}(0,0)}$ in this formula has the following important interpretation. Let $\tau=\min\{t:x(t)=0\}$. Then
\begin{equation} \label{abab}
\frac{R_{\lambda}^{(0)}(x,0)}{R_{\lambda}^{(0)}(0,0)}=E_xe^{-\lambda \tau}=\int_0^\infty e^{-\lambda s}q_x(s)ds,
\end{equation}
where $q_x(s)$ is the distribution density of the random variable $\tau$ (if the process starts at $x$.) The inverse Laplace transform of (\ref{res1ab})
leads to the identity
\begin{equation} \label{abad}
p_1(t,x,x)=p_0(t,x,x)-\int_0^t q_x(s)p_0(t-s,x,0)ds.
\end{equation}

One can derive from (\ref{resla}) that
$$
R_{\lambda}^{(0)}(x,0)\sim \frac{1}{2\pi}\ln (\sqrt\lambda |x|)~~ \text{as} \quad x\in Z\backslash\{0\};\quad R_{\lambda}^{(0)}(0,0)\sim \frac{1}{2\pi}\ln \sqrt\lambda,~~
$$
when $\sqrt\lambda |x|\to+0,~\lambda\to+0$, respectively, i.e.,
$$
\frac{R_{\lambda}^{(0)}(x,0)}{R_{\lambda}^{(0)}(0,0)}\sim \frac{\ln (\sqrt\lambda |x|)}{\ln \sqrt\lambda},~~\sqrt\lambda |x|\to+0.
$$
(The latter formula has an analog in the continuous case if the annihilation occurs at the unit circle, and $\tau$ is the time needed to reach the circle. Then the Laplace transform of the distribution density of $\tau$ equals $\frac{K_0 (\sqrt\lambda |x|)}{K_0 (\sqrt\lambda)}\sim \frac{\ln (\sqrt\lambda |x|)}{\ln \sqrt\lambda}$). The latter formula and (\ref{abab}) after the rescaling $\lambda=\lambda_1|x|^{-\alpha}$, $\alpha>2,$ imply
$$
E_xe^{-\lambda_1\frac{ \tau}{|x|^\alpha}}\rightarrow \frac{\alpha-2}{\alpha} \quad \text{as} \quad |x|\to \infty.
$$
Since the right-hand side above does not depend on $\lambda_1$, it follows that $\frac{\tau}{|x|^\alpha}$ converges in law to zero or infinity with the probabilities $\frac{\alpha-2}{\alpha}$ and $\frac{2}{\alpha}$, respectively:
$$
P_x\{\frac{\tau}{|x|^\alpha}\to 0\}=\frac{\alpha-2}{\alpha},~~P_x\{\frac{\tau}{|x|^\alpha}\to \infty\}=\frac{2}{\alpha},~~|x|\to\infty.
$$
The latter relation after the substitution $s=|x|^\alpha,~\alpha>2,$ (i.e., $\alpha=\frac{\ln s}{\ln|x|}$) leads to
\begin{equation}\label{xxx}
P_x\{\tau>s\}\sim \frac{2\ln|x|}{\ln s}~~~\text {when}~~|x|\to\infty,~~\frac{s}{|x|^2}\to\infty.
\end{equation}
After  formal differentiation (in fact, we do not need to use the differential form of the relation above), we get (compare to Lemma \ref{ogr2})
$$
q_x(s)\sim \frac{2\ln|x|}{s\ln^2s},\quad s>\gamma(x),~~|x|\to\infty.
$$
The main estimate (\ref{ababa}) follows from (\ref{xxx}), formula (\ref{abad}) and standard Gaussian estimate for $p_0$:
$$
p_0(t,x,0)=\frac{e^{-\frac{x^2}{4t}}}{4\pi t}(1+O\frac{1}{\sqrt t}),~~~|x|>t^{3/2}.
$$

\qed
\section{Estimates from below} The goal of this section is to show that the estimates (\ref{tranz})-(\ref{barg}) are sharp in the following sense: the operator has infinitely many negative eigenvalues in the case of any potential which decays at infinity a little slower (by a logarithmic factor) than in those estimates. To be more exact, the following theorem holds
\begin{theorem}Let $H=-\Delta-V(x)$ be a one-dimensional Schr\"{o}dinger operator in $L^2(Z)$ or $L^2(R)$ with the potential $V$ such that for some $\varepsilon>0,$
\[
\sum_Z \frac{ |x|}{\ln^{1+\varepsilon}(1+|x|)} V(x)=\infty, \quad \text{or} \quad \int_R \frac{ |x|}{\ln^{1+\varepsilon}(1+|x|)}V(x)dx=\infty,
\]
respectively. Then $H$ has infinitely many negative eigenvalues ($N_0(V)=\infty$).

Let $H=-\Delta-V(x)$ be a two-dimensional Schr\"{o}dinger operator in $L^2(Z^2)$ and
\begin{equation} \label{6}
\sum_{Z^2}V(x)=\infty.
\end{equation}
Then $N_0(V)=\infty$.
\end{theorem}
\textbf{Proof.} We will prove the first statement ($d=1$) only in  the lattice case. The continuous case can be treated similarly (and in fact, it is simpler). Consider sets $l=l_k=\{x:2^k\leq x\leq 2^{k+1}\}\subset Z$. Let
\begin{equation} \label{3}
a_k=\sum_{l_k} \frac{ |x|}{\ln^{1+\varepsilon}(1+|x|)} V(x), \quad k
\geq 1.
\end{equation}
Since $\sum a_k=\infty$, there exists an infinite sequence of values of $k=k_j,~j=1,2,...,$ for which
\begin{equation} \label{4}
a_k>k^{-(1+\varepsilon /2)}, \quad k=k_j.
\end{equation}
By taking a subsequence, if needed, we can guarantee that $k_{j+1}-k_j\geq 2$. Let $L_k=\{x:2^{k-1}\leq x\leq 2^{k+2}\},~ k=k_j,$ be the union of $l_{k_j}$ and two neighboring sets $l_k$. The sets $\{L_{k_j}\}$ do not have common points (except, perhaps, the end points). The first statement of the theorem will be proved if, for infinitely many sets $L=L_{k_j}$, we construct functions $\psi=\psi_j$ with the support in $L$ and such that $(H\psi,\psi)<0$.

We will take
\[
\psi=\sin[\frac{\pi}{|L|}(x-a)],~~ x\in L, \quad \psi =0,~~ x \notin L,
\]
where $|L|=2^{k+2}-2^{k-1},~k=k_j,$ is the length of the interval between the end points of $L$ and $a=2^{k_j-1}$ is the left end point of the set $L$. The function $\psi$ is a sine function whose half-period is $L$ and which is zero outside $L$. The $l^2(Z)-$norm of this function for large $L$ has order $\sqrt{L/2}$:
\[
||\psi||=\sqrt{|L|/2}(1+o(1)), \quad |L|\to \infty.
\]
One can easily show that $-\Delta \sin\alpha x=\sigma (\alpha) \sin \alpha x,~x\in Z$,  where $\alpha $ is arbitrary and $\sigma (\alpha)=2-2\cos \alpha \sim \alpha^2$ as $\alpha \to 0$. Hence,
\[
-\Delta \psi =\sigma(\frac{\pi}{|L|}) \psi-\sin \frac{\pi}{|L|}(\delta _{a}(x)-\delta _{b}(x)),
\]
where $\delta _{y}(x)$ is the delta function at the point $y$, and $a,~b$ are the left and right end points of $L$, respectively. Thus, $(-\Delta \psi,\psi)=\sigma(\frac{\pi}{|L|}) ||\psi||^2$, and therefore
\begin{equation} \label{5}
(-\Delta \psi,\psi)\sim \frac{\pi^2}{2|L|}, \quad |L| \to \infty.
\end{equation}

Let us evaluate now
\[
(V\psi,\psi)=\sum_{x\in L}V(x)\psi^2(x)\geq \sum_{x\in l}V(x)\psi^2(x).
\]
Since $\frac{ |x|}{\ln^{1+\varepsilon}(1+|x|)}\leq C2^{k}k^{-1-\varepsilon}$ on $l_k$ and $V(x)\geq 0$, from (\ref{3}) and
(\ref{4}) it follows that
\[
\sum_{x\in l_k}V(x)\geq C_12^{-k}k^{\varepsilon/2}\geq C_1|L|^{-1}\ln^{\varepsilon/2}|L|, \quad |L|\to \infty.
\]
Furthermore, $l_k$ is located far enough from the end points of $L_k$, and there exists $c>0$ such that $\psi(x)>c,~ x\in l_k.$ Hence,
\[
(V\psi,\psi)\geq \frac{C\ln^{\varepsilon/2}|L|}{|L|}, \quad |L|\to \infty.
\]
Together with (\ref{5}), this proves that $(H\psi,\psi)\leq 0$ for large enough $L$.

The proof of the one-dimensional statement of Theorem \ref{t1} is complete.

Let us prove the statement of the theorem concerning the two-dimensional operators. As in the previous case, we will construct a sequence of functions $\psi=\psi_j(x),~x\in Z^2,$ with non-intersecting finite supports and such that $(H\psi,\psi)<0$. The functions $\psi_j$ will be defined by induction as the restrictions of some functions $\phi=\phi_j $ on the Euclidian space $R^2$ onto $Z^2 \subset R^2.$ Denote by $Q_k$ squares in $R^2$ for which $|x_1|,|x_2|\leq k$. Let us define $\phi=\phi_{j_0+1}$ while assuming that functions $\phi_j,~ j\leq j_0,$ have been constructed. We choose $k$ large enough so that the supports of all the functions $\psi_j$ already defined are located strictly inside $Q_k$. We take $k=1$ to define the first function $\phi_1$. The function $\phi=\phi_{j_0+1}$ will be supported by a square layer $P=Q_{2l}\setminus Q_{k} $ with some $l\gg k$ chosen below. Thus each layer $P_j$ is split naturally in two parts, the interior part $P^{(1)}=Q_{l}\setminus Q_{k}$ and the exterior part $P^{(2)}=Q_{2l}\setminus Q_{l}$. We put $\phi=0$ outside of $P$ and $\phi=1$ on the interior part of the layer $P$. Then we split the exterior part $P^{(2)}$ into four trapezoidal regions using diagonals of the square $Q_{2l}$ and define $\phi$ to be such a linear function in each of these trapezoidal regions that $\phi=1$ on the boundary $\partial Q_{l}$ of the square $Q_{l}$ and $\phi=0$ on $\partial Q_{2l} $. Note that $\phi=0 $ on $\partial P$.

Let us estimate $(-\Delta \psi, \psi),~ x\in Z^2,$ from above. We will use notation $\partial Q_k$ for the boundary of the square $Q_k\subset R^2$, and $q$ for the union of the boundaries of the trapezoidal regions in $P \subset R^2$ constructed above. Since $-\Delta u=0$ for any linear function $u$ on $Z^2$, the support of the function $-\Delta \psi$ belongs to the set $\partial Q_{k}\bigcup\partial Q_{k+1}\bigcup q$, i.e.,
\[
|(-\Delta \psi, \psi)|\leq \sum_{\partial Q_{k}\bigcup\partial Q_{k+1}\bigcup q}|\Delta \psi|,
\]
since $0\leq\psi\leq 1$. Furthermore, $|\psi(x_1)-\psi(x_1)|\leq 1$ for each pair of neighboring points $x_1,x_2\in Z^2$, and therefore (see (\ref{lapl})) $|-\Delta \psi|\leq 4,~x\in Q_{k}\bigcup\partial Q_{k_{j}+1}$. In fact, the latter estimate holds with $2$ instead of $4$, but we do not need this improvement. A better estimates holds on $q$. Since $|\nabla \phi|\leq 1/l$, we have $|-\Delta \psi|\leq 4/l,~x\in q$. Taking into account that $|\partial Q_{k}|+|\partial Q_{k}|\leq c_1k$ and $|l|\leq c_2l$, we arrive at
\[
|(-\Delta \psi, \psi)|\leq 4c_1k+4c_2.
\]
Note that the latter estimate does not depend on $l$.

Obviously, $(V\psi,\psi)\geq \sum _{x\in P^{(1)}}V(x)$. Assumption (\ref{6}) allows us to choose $l$ such that the right-hand side of the latter inequality exceeds $4c_1k+4c_2$. Then $(H\psi,\psi)<0$ and the proof is complete.
\qed
\section{General discrete Schr\"{o}dinger operators with recurrent underlying Markov processes}

This section is devoted to a Bergmann type estimate for general lattice operators. We also will show here that shift-invariant estimates of the form (\ref{inv}) can not be valid for operators with recurrent underlying Markov processes.

Let $X$ be a countable set and let $H_0$ be a symmetric non-negative operator on $L^2(X)$ with matrix elements $h(x,y)$, i.e.
$$
H_0\psi(x)=\sum_{y\in X}h(x,y)\psi(y), \quad h(x,y)=h(y,x).
$$
It is assumed that
$$
h(x,y)\leq 0~~\text{if}~~x\neq y,\quad \sum_{y\in X}h(x,y)= 0 ;\quad h(x,x)\leq c_0 ~~~\text {for all} ~~x\in X.
$$
Obviously, operator $H_0$ can be written in the form
$$
H_0\psi(x)=\sum_{y\in X:y\neq x}h(x,y)(\psi(y)-\psi(x)).
$$
The first two conditions above guarantee the existence and uniqueness of the Markov process $x(t)$ with the generator $-H_0$ and that the operator $H_0$ is non-negative, and the last one is needed to avoid a pathological behavior of the Markov process $x(t)$. We also assume connectivity, i.e., $X$ can not be split in two disjoint non-empty sets $X_1\cup X_2$ in such a way that $h(x_1,x_2)=0$ for each $x_1\in X_1,~x_2\in X_2$. If $H_0=-\Delta$ on $Z^d$, we have $h(x,x)=2d,~h(x,y)=-1$ when $|x-y|=1, ~h(x,y)=0$ when $|x-y|>1$.

Let $p_0(t,x,y)$ be the transition probability, i.e., $p_0$ is the kernel of the Markov semigroup $e^{-tH_0}$, and let
\begin{equation}\label{rp}
R^{(0)}_\lambda(x,y)=-\int_0^\infty  p_0(t,x,y)e^{-\lambda t}dt
\end{equation}
be the kernel of the resolvent $R^{(0)}_\lambda=(-H_0-\lambda)^{-1}$ of the operator $-H_0$. The connectivity assumption implies that $p_0(t,x,y)>0$ and $R_\lambda<0$ for all values of the arguments. Since $-H_0\leq 0,$ the function $R^{(0)}_\lambda(x,y)$ is analytic in $\lambda\notin (-\infty,0]$. We assume that the process $x(t)$ is recurrent, i.e.,
\begin{equation}\label{recu}
\int_0^\infty p_0(t,x,x)dt=\infty.
\end{equation}
The latter relation implies that
\begin{equation}\label{rinf}
\lim_{\lambda \to +0}R^{(0)}_\lambda(x,x)=-\infty.
\end{equation}

The following result is a Bargmann type estimate for the lattice operator $H=H_0-V(x)$. Let us fix an arbitrary point $x_0\in X$. Denote
$$
\widetilde{R}=\widetilde{R}(x,x_0)=\lim_{\lambda\to +0}[\frac{[R^{(0)}_\lambda(x,x_0)]^2}{R^{(0)}_\lambda(x_0,x_0)}-R^{(0)}_\lambda(x,x)]
$$
and
\begin{equation}\label{renorm}
\widetilde{R}=\widetilde{R}(x,x_0)=2\lim_{\lambda\to +0}[R_{\lambda}^{(0)}(x,x_0)-R_{\lambda}^{(0)}(x_0,x_0)]
\end{equation}
if the operator $H_0$ is translation-invariant.
\begin{theorem}\label{barggen}
Let the Markov process $x(t)$ with the generator $-H_0$ be recurrent. Then

1) the function $\widetilde{R}$
is finite for all $x,x_0\in X$ and positive for $x\neq x_0$ (it vanishes if $x=x_0$),

2) the following two estimates hold
\begin{equation}\label{f1}
N_0(V)\leq \sum_{x\in X}V(x)\widetilde{R}(x,x_0)+1,
\end{equation}
\begin{equation}\label{f2}
N_0(V)\leq \#\{x\in X:~V(x)\geq 1\}+\sum_{x:V(x)<1}V(x)\widetilde{R}(x,x_0)+1.
\end{equation}
\end{theorem}
\textbf{Proof.} Let us denote by $\widetilde{x}(t)$ the Markov process $x(t)$ with the additional condition of annihilation at the point $x_0\in X.$ From the connectivity assumption it follows that the conditional process  $\widetilde{x}(t)$ with the annihilation at $x_0\in X$ is transient. With this fact taken into account, the above formulas for $\widetilde{R}$ differ from (\ref{res}), (\ref{res1}) only by sign and by $x_0$ playing the role of the origin $x=0$. Thus,
$$
\widetilde{R}(x,x_0)=-R^{(1)}_\lambda(x,x)=\int_0^\infty p_1(t,x,x)dt,
$$
where $p_1(t,x,y)$ is the transition probability for the process $\widetilde{x}(t)$. This implies the first statement of the theorem. The latter relation and (\ref{1xxa}) with $\sigma=0$ lead to (\ref{f1}). In order to obtain  (\ref{f2}), one can consider the potential $\widetilde{V}$ which is obtained by reducing the values of $V$ to zero at all points where $V\geq 1$. Then  (\ref{f2}) follows from  (\ref{f1}) for the potential  $\widetilde{V}$ since the operators with these two potentials differ by an operator of the rank $N=\#\{x\in X:~V(x)\geq 1\}$.
\qed

Theorem \ref{tno} below shows that a space invariant estimate of $N_0(V)$ can not be valid for the discrete operator $H$, but first we need to prove the following preliminary result.
\begin{theorem}\label{tinter}
For each $y\in X$ and $\alpha>0,$ the operator $H=H_0-\alpha\delta_{y}(x)$ has a unique simple negative eigenvalue $\lambda=\lambda (y,\alpha)<0$.
\end{theorem}
\textbf{Proof.} The uniqueness is due to the fact that $H$ is a rank one perturbation of  $H_0$. Let us show the existence of the eigenvalue. First we note that $\sum_x p_0(t,x,y)=1$, and therefore (\ref{rp}) implies that
\begin{equation}\label{sum1}
 \sum_x R^{(0)}_\lambda(x,y)=\frac{-1}{\lambda}, \quad\lambda>0.
\end{equation}
Formula (\ref{rp}) implies also that $R^{(0)}_\lambda(x,y)< 0$. Thus from (\ref{sum1}) it follows that $|R^{(0)}_\lambda(x,y)|\leq \frac{1}{\lambda}$ for each $x,y\in X$ and $\lambda>0$. This and (\ref{sum1}) leads to the estimate $\sum_x [R^{(0)}_\lambda(x,y)]^2\leq \frac{1}{\lambda^2},$ i.e., $R^{(0)}_\lambda(x,y)\in L^2(X),~y\in X$.

We look for an eigenfunction in the form $\psi_\lambda(x)=R^{(0)}_\lambda(x,y),~\lambda>0$. Since $(-H_0-\lambda)\psi_\lambda=\delta_{y}(x)$, $\psi_\lambda$ will be an eigenfunction of $H=H_0-\alpha\delta_{y}(x)$ with the eigenvalue $-\lambda$ if $-\alpha R^{(0)}_\lambda(y,y)=1$. The latter equation has a solution $\lambda=\lambda (y,\alpha)>0$ for each $y\in X$ and $\alpha>0$ due to (\ref{rinf}) and the relation $\lim_{\lambda\to \infty} R^{(0)}_\lambda(y,y)=0.$
\qed

The next theorem shows that for each $\gamma$ one can find a potential $V\geq 0$ such that $\sum_{x\in X} V^\gamma (x)$ is arbitrary small and the operator $H$ has infinitely many negative eigenvalues. Hence, estimate (\ref{inv}) can not be valid for the operator $H$. The potential $V$ will be constructed when a uniformity condition on the unperturbed operator $H_0$ holds. We assume that there exists an integer-valued metric $d(x,y)$ on $X$ (for example, $l^1-$metric on $Z^d$) such that the following two relations hold:

a) $|R^{(0)}_\lambda(x,x)|\geq \beta(\lambda)$ for $\lambda>0$ and some $\beta(\lambda)>0$, and $\beta(\lambda)\to \infty$ as $\lambda\to+0$;

b) $\sum_{x:d(x,y)>r}| R^{(0)}_\lambda(x,y)|$ tends to zero uniformly in $y$ when $\lambda>0$ and $r\to\infty.$

\begin{theorem}\label{tno}
Let conditions a),b) hold. Then for any sequence $\alpha_n\to+0,$ one can find a set of points $\{x_n\in X\}$ such that the operator
\begin{equation}\label{fullop}
H=H_0-\sum_{n=1}^\infty\alpha_n\delta_{x_n}(x)
\end{equation}
has infinitely many negative eigenvalues.
\end{theorem}
\textbf{Proof.} In order to prove the theorem, it is sufficient to construct a sequence of compactly supported functions $\{\psi_k(x)\}$ with disjoint supports such that
\begin{equation}\label{diri}
(H\psi_k(x),\psi_k(x))<0.
\end{equation}

For fixed $y\in X,\alpha>0$, consider a ``test" operator
$$
H=H(y,\alpha)=H_0-\alpha\delta_{y}(x).
$$
Due to the previous theorem, this operator has a negative eigenvalue $-\lambda_0(y,\alpha)$, where $\lambda=\lambda_0(y,\alpha)>0$ is the root of the equation  $-\alpha R^{(0)}_\lambda(y,y)=1$. The corresponding eigenfunction can be chosen as
$$
\psi(x)=\frac{R^{(0)}_{\lambda_0}(x,y)}{\sqrt{\sum_x[R^{(0)}_{\lambda_0}(x,y)]^2}}.
$$
Note that condition a) implies that $\lambda_0\geq \lambda_0(\alpha)>0$, where the lower bound $\lambda_0(\alpha)$ does not depend on $y$.

In order to complete the proof of the theorem, we will need the following lemma.
\begin{lemma}\label{unif}
There exists a function $r=r(\alpha)$ such that the inequality
\begin{equation} \label{un1}
(H\widetilde{\psi}(x),\widetilde{\psi}(x))<0,~~~H=H(y,\alpha),
\end{equation}
holds for the truncated eigenfunction
$$
\widetilde{\psi}(x)=\psi(x)I_{d(x,y)\leq r(\alpha)}.
$$
\end{lemma}

The important part of the statement of this lemma is that $r$ is $y-$independent. The statement follows from the uniformity assumption. Indeed,
$$
(H\psi(x),\psi(x))=-\lambda_0(y,\alpha)\leq-\lambda_0(\alpha)<0.
$$
Hence, it is enough to show that
$$
|(H\psi,\psi)-(H\widetilde{\psi},\widetilde{\psi})|=
|(H\psi,\psi-\widetilde{\psi})+(H(\psi-\widetilde{\psi}),\widetilde{\psi})|<\frac{\lambda_0(\alpha)}{2}\quad \text{when}~~r>r(\alpha).
$$
Since the operator $H$ is bounded in  $l^2(X)$, $\|\psi\|=1,~\|\widetilde{\psi}\|\leq 1$, it remains to prove that $\|\psi-\widetilde{\psi}\|\to 0$ uniformly in $y$ when $\alpha$ is fixed and $r\to\infty$ (all the norms here and below are in $l^2(X)$). It was shown in the proof of Theorem \ref{tinter} that $|R^{(0)}_\lambda(x,y)|\leq \frac{1}{\lambda}$ for each $x,y\in X$ and $\lambda>0$. Thus, from condition b) and the estimate $\lambda_0(y,\alpha)\geq \lambda_0(\alpha)>0$ it follows that
$$
\|(1-I_{d(x,y)\leq r})R^{(0)}_{\lambda_0}(x,y)\|^2\leq \frac{1}{\lambda_0}\sum_{x:d(x,y)>r}| R^{(0)}_{\lambda_0}(x,y)| \to 0
$$
uniformly in $y$ when $r\to\infty.$ Condition a) implies that
$$
\sum_x[R^{(0)}_{\lambda_0}(x,y)]^2\geq [R^{(0)}_{\lambda_0}(y,y)]^2\geq \beta(\lambda_0)>0.
$$
This completes the proof of the lemma since
$$
\|\psi-\widetilde{\psi}\|^2=
\frac{\|(1-I_{d(x,y)\leq r})R^{(0)}_{\lambda_0}(x,y)\|^2}{\sum_x[R^{(0)}_{\lambda_0}(x,y)]^2}.
$$

Let us complete the proof of the theorem.
We fix $\alpha_1$, calculate $r=r(\alpha_1)$, select an arbitrary point $x_1$ and chose $\psi_1(x)$ to be the truncated eigenfunction of the ``test" operator $H(x_1,\alpha_1)$. Other points $x_n,~n>1,$ will be chosen outside of the support of $\psi_1.$ Thus inequality (\ref{un1}) with the ``test" operator $H(x_1,\alpha_1)$ implies the same inequality for operator (\ref{fullop}), i.e., (\ref{diri}) holds for $\psi_1$. Now we fix $\alpha_2$, calculate $r=r(\alpha_2)$, select an arbitrary point $x_2$ such that $d(x_2,x_1)>r(\alpha_2)+r(\alpha_1)$, and chose $\psi_2(x)$ to be the truncated eigenfunction of the ``test" operator $H(x_2,\alpha_2)$. The supports of functions $\psi_1$ and $\psi_2$ are disjoint, and other points $x_n, ~n>2,$ will be chosen outside of the supports of $\psi_1,\psi_2$. Hence,  (\ref{diri}) holds for $\psi_2$. The point $x_3$ will be chosen in such a way that  $d(x_3,x_1)>r(\alpha_3)+r(\alpha_1)$ and $d(x_3,x_2)>r(\alpha_3)+r(\alpha_2)$, etc.. This procedure allows us to construct the desired sequence $\{\psi_k(x)\}$.
\qed

\section{Fractional power of the lattice operator}

This section provides an illustration of the results on general discrete operators obtained in the previous section. It concerns an important class of non-local random walks  $x_\alpha(t)$ on the one-dimensional lattice. The one-dimensional lattice Laplacian
$$
-H_0 =\Delta\psi(x)=\psi(x+1)+\psi(x-1)-2\psi(x)
$$
on $l^2(Z)$ is the generator of the symmetric random walk $x(t)$ with continuous time. Let $P_t=e^{t\Delta}$ be the corresponding semigroup, $P_t\psi=\sum_{y\in Z}p(t,x,y)\psi(y)$. The operator
$\Delta$ in the Fourier space $L^2[-\pi,\pi]$ acts as the multiplication by the symbol
$$
\widehat{\Delta}(\phi)=2(\cos\phi-1)=-4\sin^2\frac{\phi}{2},~~\phi\in[-\pi,\pi].
$$
Similarly,
$$
\widehat{P_t}(\phi)=e^{-4t\sin^2\frac{\phi}{2}}, \quad \widehat{R}_\lambda(\phi)=-\int_0^\infty e^{-\lambda t}\widehat{P_t}dt=\frac{-1}{\lambda+4\sin^2\frac{\phi}{2}},~~\phi\in[-\pi,\pi].
$$

The main object that we study in this section is the fractional degrees $H_0^\alpha, ~\alpha > 0,$ of the operator $H_0=-\Delta$. After the Fourier transform, the operator $H_0^\alpha$ , its semigroup and the resolvent are the operators of multiplication by the symbols
$$
\widehat{(-\Delta)^\alpha}=\left(4\sin^2\frac{\phi}{2}\right)^\alpha, \quad \widehat{P_{t,\alpha}}=e^{-t\left(4\sin^2\frac{\phi}{2}\right)^\alpha}, \quad
\widehat{R_{\lambda,\alpha}}=\frac{-1}{\lambda+(4\sin^2\frac{\phi}{2})^\alpha}.
$$

The following result is well-known in probability theory.
\begin{lemma}\label{polo}
The operator $-H_0^\alpha=-(-\Delta)^\alpha,~\alpha>0,$ is the generator of a Markov process $x_\alpha(t)$ on $Z$ iff $0< \alpha \leq 1$.
\end{lemma}
\textbf{Proof.} One needs only to prove the positivity of the kernel $p_\alpha (t,x,y)$ of the semigroup $P_{t,\alpha}$ for $0<\alpha<1$ and non-positivity of the kernel for $\alpha>1$. If $0<\alpha< 1$, then there exists \cite{F} (Ch. 13, 6) the probability density $g_{\alpha, 1}(s)>0,~0< s<\infty,$ (which is called the stable law with the parameters $\alpha$ and $\beta=1$) such that
$$
e^{-\lambda^\alpha}=\int_0^\infty e^{-\lambda t}g_{\alpha, 1}(t)dt.
$$
The second parameter $\beta$ in the two-parametric family of the densities $g_{\alpha, \beta}$ characterizes the symmetry of the density. If $\beta=0$ then $g_{\alpha, 0}(s)= g_{\alpha, 0}(-s)$, if $\beta=1,~0< \alpha <1,$ then $g_{\alpha, 1}(s)=0,~s\leq 0,~ g_{\alpha, 1}(s)>0,~s>0$.

Using the probability density $g_{\alpha, 1}$, one can represent $P_{t,\alpha}$ in the form
$$
P_{t,\alpha}=e^{-t(-\Delta)^\alpha}=\int_0^\infty e^{t^{1/\alpha} s\Delta}g_{\alpha, 1}(s)ds=\int_0^\infty P_{t^{1/\alpha}s}g_{\alpha, 1}(s)ds,
$$
i.e., the kernels $p_\alpha$ and $p$ of the operators $P_{t,\alpha},~P_t$ are related by
$$
p_\alpha(t,x,y)=\int_0^\infty p(t^{1/\alpha} s,x,y)g_{\alpha, 1}(s)ds.
$$
This implies the positivity of $p_\alpha$.

In order to show that the semigroup $P_{t,\alpha}$ is not positive when $\alpha>1$, we note that the function
$
\widehat{h}(t,\phi)=\widehat{P_{t,\alpha}}(\phi)
$ has the following property: $\widehat{h}''(0)=0$. Hence, its inverse Fourier transform $h(t,z)=\frac{1}{2\pi}\int_{-\pi}^\pi \widehat{h}(t,\phi)e^{iz\phi}d\phi$ satisfies $\sum_{z\in Z}z^2h(t,z)=0$, which shows that $p_\alpha=h(t,x-y)$ can not be non-negative.
\qed

\begin{lemma}\label{polo}
For each $\alpha\in (0,1]$ and $t\to\infty$,
$$
p_\alpha(t,x,x)\sim \frac{c_\alpha}{t^{1/(2\alpha)}},  \quad  c_\alpha=\frac{\Gamma(1/(2\alpha))}{2\pi\alpha}.
$$
\end{lemma}
\textbf {Corollary.} The random walk $x_\alpha(t)$ is transient for $0<\alpha<1/2$ and recurrent for $1/2\leq\alpha\leq 1.$ The formula above indicates that $p_\alpha(t,x,x)$ has the same asymptotic behavior as the transition probability for ``nearest neighbors random walks" (defined by the standard Laplacian) when the dimension $d$ equals $1/\alpha$. A similarity between the long range 1-D ferromagnetic interaction and high-dimensional local interaction (similar to noted above) was discovered by Dyson \cite{Dy}. This similarity was a foundation for the introduction of the hierarchical lattice. We will discuss the spectral properties of the hierarchical Dyson's Laplacian elsewhere.

\textbf {Proof.} This statement is a simple consequence of the Laplace method applied to the integral
$$
p_\alpha(t,x,x)=\frac{1}{2\pi}\int_{-\pi}^\pi e^{-t\left(4\sin^2\frac{\phi}{2}\right)^\alpha}d\phi\sim \frac{1}{2\pi}\int_{-\pi}^\pi e^{-t|\phi|^{2\alpha}}d\phi, \quad t\to\infty.
$$
\qed

\textbf{Remark.} Similar calculations give a more general result. If $x_\alpha(t)$ is a random walk on $Z$ with the generator $-(-\Delta)^\alpha$, then
\begin{equation} \label{law}
\frac{x_\alpha(t)}{t^{1/\alpha}}\rightarrow \phi \quad \text{in law as} ~~t\to\infty,
\end{equation}
where $\phi$ has a stable distribution $g_{2\alpha,0}(s),~s\in R,$ (symmetric stable law with parameters $2\alpha, \beta=0$ and characteristic function (Fourier transform) equal to $e^{-\lambda^{2\alpha}}$). Indeed,
$$
E_0e^{i\frac{\lambda x_\alpha(t)}{t^{1/\alpha}}}=e^{-t\left(4\sin^2\frac{\lambda}{t^{1/\alpha}}\right)^\alpha}\to e^{-\lambda^{2\alpha}},~~~t\to\infty.
$$
Formula (\ref{law}) means that, after rescalling, the lattice operator $(-\Delta)^\alpha$ approximates the fractional power of the continuous Laplacian, i.e., the random walk $x'_\alpha(s)=x_\alpha(st)/t^{1/\alpha}$ (which is the rescalling of $x_\alpha(s)$) converges to the stable process $x^*_\alpha(s)$ on $R$ with the generator $-(-\frac{d^2}{dx^2})^\alpha$.

The following theorem is an immediate consequence of the standard CLR estimate (\ref{1xx}) with $\sigma=0$, Lemma \ref{polo} and finite rank perturbation arguments (see Remark 2 after Theorem \ref{t42})

\begin{theorem} (Transient case) Consider the Hamiltonian
$$
H_\alpha=-(-\Delta)^\alpha-V(x) \quad \text{on} ~~l^2(Z),~~~V(x)\geq 0,
$$
with $0<\alpha<1/2$ (i.e., the dimension $d=\frac{1}{\alpha}>2$). Then there is a constant $C=C(\alpha)$ such that
$$
N_0(V)\leq \#\{x\in Z: V(x)\geq 1\}+C(\alpha)\sum_{x:V(x)<1}V^{\frac{1}{2\alpha}}(x).
$$
\end{theorem}
\textbf{Remark.} The constant $C(\alpha)$ can be evaluated. One can show that $C(\alpha)=O(\frac{1}{1-2\alpha})$ as $\alpha\to 1/2.$

Let us consider now the recurrent case: $\alpha\geq 1/2.$ First we calculate the regularized resolvent (\ref{renorm}):
$$
\widetilde{R}_0(x,0)=\lim_{\lambda\to +0}[R_{\lambda,\alpha}(x,0)-R_{\lambda,\alpha}(0,0)]=\lim_{\lambda\to +0}
\frac{1}{2\pi}\int_{-\pi}^\pi \frac{1-e^{i\phi x}}{\lambda+(4\sin^2\frac{\phi}{2})^\alpha}d\phi
$$
$$
=\lim_{\lambda\to +0}\frac{2}{\pi}\int_{0}^\pi \frac{\sin^2(\frac{\phi}{2}x)}{\lambda+(4\sin^2\frac{\phi}{2})^\alpha}d\phi
=\frac{2}{\pi}\int_{0}^\pi \frac{\sin^2(\frac{\phi}{2}x)}{(4\sin^2\frac{\phi}{2})^\alpha}d\phi.
$$

A simple analysis provides the following asymptotics of the regularized resolvent when $|x|\to \infty:$ if $\alpha>1/2$, then
$$
\widetilde{R}_0(x,0)\sim c_\alpha|x|^{2\alpha-1}, \quad c_\alpha=\frac{4}{\pi}\int_0^\infty\frac{\sin^2z}{z^{2\alpha}}dz=O(\frac{1}{2\alpha-1}) \quad \text{as} ~~\alpha\to 1/2.
$$
If $\alpha=1/2$ (a borderline case which corresponds to a Cauchy-type random walk with the generator $-(-\Delta)^{1/2}$), then
$$
\widetilde{R}_0(x,0)\sim \frac{1}{\pi}\ln|x|.
$$
Hence, Theorem \ref{barggen} implies
\begin{theorem} (Recurrent process)
There exist constants $C_\alpha, C$ such that
$$
N_0(V)\leq \#\{x\in R:~V(x)\geq 1\}+C_\alpha\sum_{x:V(x)<1}V(x)|x|^{2\alpha-1}+1, \quad ~~\frac{1}{2}<\alpha<1,
$$
$$
N_0(V)\leq \#\{x\in R:~V(x)\geq 1\}+C\sum_{x:V(x)<1}V(x)\ln(2+|x|)+1, \quad ~~\alpha=1/2.
$$
\end{theorem}
\section{Bessel operators}

This section concerns another class of one-dimensional operators which may have an arbitrary positive spectral dimension. These operators $B_d$ on the half line $R_+$ are defined by the radial part of the Laplacian:
$$
B_d=\frac{d^2}{dr^2}+\frac{d-1}{r}\frac{d}{dr},\quad r>0.
$$
We consider arbitrary (not necessarily integer) $d>0$. The operator $B_d$ can be represented in the form
\begin{equation} \label{bessb}
B_d=\frac{1}{r^{d-1}}\frac{d-1}{r}(r^{d-1}\frac{d}{dr}), \quad d> 0,
\end{equation}
i.e., it is symmetric in $L^2([0,\infty), r^{d-1}dr)$.

If $d\geq 2$, then the operator $B_d$ in  $L^2([0,\infty), r^{d-1}dr)$ is self-adjoint (we do not need to impose a boundary condition at $r=0$). The diffusion process $b_d(t)$ with the generator $B_d$ is also well-defined since the point $r=0$ is not accessible from any initial point $r>0$. If $d<2$, the situation is different. The equation $B_d\psi=0$ has two bounded linearly independent solutions, $\psi_1=1, \psi_2=r^{2-d}$, i.e., due to the Weil criterion, we have the limit circle case near $r=0,$ and a boundary condition is needed to define a self-adjoint operator. One can impose the Dirichlet boundary condition at $r=0$ which corresponds to annihilation of the process  $b_d(t)$ at $r=0$. Another option is an analog of the classical Neumann boundary condition: $\lim_{r\to +0}r^{d-1}\psi'(r)=0$ (see \cite{Ito}).

We will consider the Schr\"{o}dinger operator $H_d=-B_d-V(r)$ with the Dirichlet boundary condition at $r=0$ if $d<2$ and without a boundary condition if $d\geq 2$.  The process $b_d(t)$ in both cases is transient, and our main concern is to obtain an exact formula for the transition probability $p_d(t,a,r)$ with respect to measure $d\mu=r^{d-1}dr.$ Denote by $I_\nu$ the modified Bessel function of order $\nu$.
\begin{lemma}\label{lbes}
The process $b_d(t)$ has the following transition density if $d\geq 2$ or $d<2$ and the Neumann boundary condition at $r=0$ is imposed:
$$
p_d(t,a,r)=(2t)^{-1}e^{-\frac{a^2+r^2}{4t}}(ar)^{1-d/2}I_{d/2-1}(\frac{ar}{2t}).
$$
If $d<2$ and the Dirichlet  boundary condition is imposed, then
$$
p_d(t,a,r)=p^D_d(t,a,r)=(2t)^{-1}e^{-\frac{a^2+r^2}{4t}}(ar)^{1-d/2}I_{1-d/2}(\frac{ar}{2t}).
$$
\end{lemma}

The first formula can be found in \cite{Ito}. The second formula can be proved similarly.

From Lemma \ref{lbes} it follows that
$$
p_d(t,r,r)\sim \frac{c_d}{t^{d/2}} \quad t\to\infty,\quad d\geq 2,~\quad p^D_d(t,r,r)\sim c_d\frac{r^{4-d}}{t^{2-d/2}} \quad t\to\infty,\quad d< 2.
$$
Applying the CLR estimate, we obtain
\begin{theorem}
If $d>2$, then
$$
N_0(V)\leq c(d)\int_0^\infty V^{d/2}r^{d-1}dr.
$$
If  $d<2$ and the Dirichlet boundary condition at $r=0$ is imposed, then
$$
N^D_0(V)\leq c_1(\sigma)\int_{r:r^2V>\sigma}V(r)r^{2-d}dr+c_2(\sigma)\int_{r:r^2V<\sigma}V^{2-d/2}(r)r^{4-2d}dr.
$$
\end{theorem}
The standard rank one perturbation arguments imply that the last estimate with constant one added to right-hand side is valid for $N^N_0(V)$.

\section{Lieb-Thirring sums}

The results of this section are based on two known formulas for the Lieb-Thirring sums for the general  Schr\"{o}dinger operators $H=H_0-V(x)$ on a complete $\sigma$-compact metric space $X$. The first formula is valid under the same assumptions, that are needed for formula (\ref{1xx}) (in particular, the transience of the underlying Markov process is required), and has the form
\begin{equation} \label{lit}
S_\gamma(V)\leq\frac{1}{c(\sigma)}\int_X V^{1+\gamma}(x)\int_{\frac{\sigma}{V(x)}}^\infty p_0(t,x,x)dt\mu(dx).
\end{equation}
Note that formula (\ref{lith}) for the operator  $H=-\Delta-V(x)$ in $R^d,~d\geq 3$, is an immediate consequence of (\ref{lit}). The second formula is valid under the same conditions, but the transience requirment is replaced by the convergence of the following integral:
$$
\int_1^\infty t^{-\gamma}p_0(t,x,x)dt<\infty.
$$
If the latter integral converges, then
\begin{equation} \label{lithi}
S_\gamma(V)\leq\frac{2\gamma \Gamma(\gamma)}{c(\sigma)}\int_X V(x)\int_{\frac{\sigma}{V(x)}}^\infty t^{-\gamma}p_0(t,x,x)dt\mu(dx),
\end{equation}
where $\Gamma(\gamma)$ is the Gamma-function. Note that (\ref{lithi}) implies (\ref{lith}) for the operator  $H=-\Delta-V(x)$ in $R^d$ when $\frac{d}{2}+\gamma>1,$ i.e., the case $d=1,~ \gamma\leq 1/2$  is still not covered by (\ref{lit}),(\ref{lithi}). Estimates for this case will be obtained below. Our approach allows us also to obtain new estimates in the cases when  (\ref{lit}) or (\ref{lithi}) hold, and in some cases these new estimates are better than  (\ref{lit}) or (\ref{lithi}). For example, our estimates on $S_\gamma(V)$ for the two-dimensional Schr\"{o}dinger operators are uniform in $\gamma\in[0,1]$, while the right-hand side in
 (\ref{lithi}) goes to infinity for these operators when $\gamma\to 0.$

 While estimate (\ref{lit}) can be found in many papers starting from the original paper by Lieb and Thirring \cite{Lt1} (see also \cite {rs}, \cite{1}, \cite{mv}), we didn't find a reference for  (\ref{lithi}) (similar formulas can be found in \cite{rs}). Thus we decided to give a brief proof of it. Let $N_E(V)=\#\{\lambda_i\leq -E,E>0\}$. We have
$$
S_\gamma(V)=\gamma\int_0^\infty E^{\gamma-1}N_E(V)dE\leq 2\gamma \int_0^\infty E^{\gamma-1}\text{Tr}(V[(H_0+E)^{-1}-(H_0+V+E)^{-1}])dE
$$
$$
\leq 2\gamma \int_0^\infty E^{\gamma-1}\text{Tr}(V\int_0^\infty[e^{-t(H_0+E)}-e^{-t(H_0+V+E)}]dt)dE
$$
$$
=2\gamma \Gamma(\gamma)\int_0^\infty t^{-\gamma}\text{Tr}(V[e^{-tH_0}-e^{-t(H_0+V)}])dt.
$$
The right-hand sides here and in (\ref{lithi}) coincide, and this justifies (\ref{lithi}).

Consider now a Schr\"{o}dinger operator $H=H_0-V(x)$ on a metric space $X$ such that the Markov process $x(t)$, generated by $-H_0$, is recurrent, and a point $x_0$
is accessible from any initial point. Then the process $x_1(t)$ with annihilation at the moment of the first arrival to $x_0$ is transient. For example, $H_0$ can be a negative lattice Laplacian on $Z^d,d\leq 2,$ the general discrete operator discussed in section 6, or the generator of a 1-D diffusion process, say, $H_0=-\frac{d^2}{dx^2}, ~x\in R,$ or $H_0=-B_d, d<2$, see (\ref{bessb}). Let $H_1$ be the generator of the process $x_1(t)$. It is given by $H_0$ with the Dirichlet boundary condition at $x_0:~\psi(x_0)=0$. Let $p_1(t,x,y)$ be the transition probability for the process $x_1(t)$.

We will assume additionally that the potential is bounded: $V(x)\leq \Lambda.$ This implies that the ground state $\lambda_0(V)$ is bounded from below, $\lambda_0(V)\geq -\Lambda$. Since the operator $\widetilde{H}=H_1-V(x)$ is a rank one perturbation of $H=H_0-V(x)$, the eigenvalues of the operators $\widetilde{H}$ and $H$ alternate. Hence, the bound for the ground state and estimates (\ref{lit}),(\ref{lithi}) applied to the operator $\widetilde{H}$ lead to the following statement.
\begin{theorem}\label{tlt}
Let $S_\gamma(V)$ be the Lieb-Thirring sum for the Schr\"{o}dinger operator $H=H_0-V(x)$, where $H_0$ is an operator which satisfies the conditions described above, and $V(x)\leq \Lambda$. Then the following two estimates hold
\begin{equation} \label{lit9}
S_\gamma(V)\leq\Lambda^\gamma+\frac{1}{c(\sigma)}\int_X V^{1+\gamma}(x)\int_{\frac{\sigma}{V(x)}}^\infty p_1(t,x,x)dt\mu(dx),
\end{equation}
\begin{equation} \label{lithi9}
S_\gamma(V)\leq\Lambda^\gamma+\frac{2\gamma \Gamma(\gamma)}{c(\sigma)}\int_X V(x)\int_{\frac{\sigma}{V(x)}}^\infty t^{-\gamma}p_1(t,x,x)dt\mu(dx),
\end{equation}
\end{theorem}
\textbf{Remark.} The second formula can be applied to the potentials which decay at infinity slower than in (\ref{lit9}), but it worsens when $\gamma\to\gamma_0$ and $t^{-\gamma_0}p_1$
 is not integrable at zero.

 Let us apply the latter theorem to the one-dimensional Schr\"{o}dinger operator on $R$. We choose $x_0=0.$ Using formula (\ref{lit9}) we arrive at (compare with (\ref{rebarg}))
\begin{theorem}\label{25}
Let $H=-
\frac{d^2}{dx^2}-V(x)$ be the one-dimensional Schr\"{o}dinger operator on $L^2(R)$ and $0\leq V(x)\leq \Lambda$. Then
\begin{equation} \label{1to1}
S_\gamma(V)\leq \Lambda^\gamma+\frac{1}{c(\sigma )}[\int_{x^2V(x)>\sigma} |x| V^{1+\gamma}(x)dx+\frac{1}{\sqrt{\sigma\pi}}\int_{x^2V(x)<\sigma} x^2 V^{3/2+\gamma}(x)dx].
\end{equation}
\end{theorem}

Using (\ref{lithi9}) and the same arguments as in section 2 we obtain the following statement.
\begin{theorem}\label{tlt}
Let $H=-
\frac{d^2}{dx^2}-V(x)$ be the one-dimensional Schr\"{o}dinger operator on $L^2(R)$ and $0\leq V(x)\leq \Lambda$. Then for any $\gamma<1/2$,
$$
S_\gamma(V)\leq \Lambda^\gamma+\frac{1}{c(\sigma )}[c_1\int_{x^2V(x)>\sigma} |x|^{1-2\gamma} V(x)dx+c_2\int_{x^2V(x)<\sigma} x^{2} V^{3/2+\gamma}(x)dx],
$$
where
$$
c_1=\frac{\gamma\Gamma(\gamma)}{\sqrt\pi}\int_0^\infty\frac{1-e^{\frac{-1}{s}}}{s^{(1+2\gamma)/2}}ds, \quad c_2=\frac{2}{(1+2\gamma)\sqrt{\sigma^{1+2\gamma}\pi}}.
$$
\end{theorem}

Note that $c_1\to \infty $ when $\gamma\to 1/2$.

Let us turn now to the two-dimensional Schr\"{o}dinger operator on $L^2(R^2)$. We can not use Theorem \ref{tlt} in this case since each point $x_0\in R^2$ is not accessible for the two-dimensional Brownian motion. Formula (\ref{lithi}) provides an estimate for $S_\gamma(V)$ with a constant which blows up when $\gamma\to 0$. One can obtain a better estimate for small $\gamma$ using annihilation due to a compactly supported potential $g(x)$ introduced in section 3: $q=1$ for $|x|<1,~q=0$ for $|x|\geq 1$. The main theorem is a consequence of (\ref{lithi}), where the operator $H=-\Delta-V(x)$ is considered as the perturbation of $-\Delta+q$ by the potential $-V-q.$

We will assume that the potential $V$ is bounded (otherwise the formula is too cumbersome). Then one can use the scaling and reduce the problem to the case when $V(x)\leq 1$.
\begin{theorem}\label{tlt}
Let $H=-
\Delta-V(x)$ in $L^2(R^2)$ and $0\leq V(x)\leq 1$. Then there exist constants $a_1,a_2<\infty$ such that for each $\gamma\in[0,1]$,
$$
S_\gamma(V)\leq a_1+a_2\int_{R^2}\frac{V^{1+\gamma}(x)}{\ln\frac{4}{V(x)}}\ln^2(2+|x|)dx.
$$
\end{theorem}
\textbf{Proof.} Let $H_1=-\Delta+q(x)$ and let $p_1(t,x,y)$ satisfies
$$
\frac{\partial p_1}{\partial t}+H_1p_1=0,~~t>0, ~~p_1(0,x,y)=\delta_y(x).
$$
Then
$$
p_1(t,x,x)\leq c_0\frac{\ln^2(2+|x|)}{t\ln^2t}, ~~t\geq2,~~~p_1(t,x,x)\leq \frac{1}{4\pi t}, ~~0<t<2.
$$
Indeed, the second estimate is due to the fact that $p_1\leq p_0$, and it is valid for all the values of the arguments. The first estimate was proved in Theorem \ref{thp1} (see also Lemma \ref{ogr3}) for $t>|x|^2\ln(2+|x|)$. It remains to note that the validity of the first estimate for $2<t<|x|^2\ln(2+|x|)$ follows immediately from the fact that $\frac{\ln^2(2+|x|)}{\ln^2t}<C<\infty$ for those $t$.

We apply formula (\ref{lithi}) with $\sigma=4$ for the operator $-\Delta+q(x)$ perturbed by the potential $-V-q$ and arrive at
$$
S_\gamma(V)\leq\frac{2\gamma \Gamma(\gamma)}{c(4)}\int_{R^2} (V(x)+q(x))\int_{\frac{4}{V(x)+q(x)}}^\infty t^{-\gamma}p_1(t,x,x)dtdx.
$$

Since
$$
A=\frac{4}{V(x)+q(x)}>2,
$$
we have
$$
\int_{A}^\infty t^{-\gamma}p_1(t,x,x)dt\leq c_0 \int_A^\infty \frac {\ln^2(2+|x|)}{t^{1+\gamma} \ln^2t}dt
$$
$$
\leq \frac{c_0\ln^2(2+|x|)}{A^\gamma}\int_A^\infty \frac {dt}{t \ln^2t}=\frac{c_0\ln^2(2+|x|)}{A^\gamma \ln A}.
$$
Thus
\begin{equation} \label{f4}
S_\gamma(V)\leq C_1\int_{R^2} \ln^2(2+|x|)G(V(x)+q(x))dx, \quad \text{where} ~~G(z)=\frac{z^{1+\gamma}}{\ln\frac{4}{z}},~~0\leq z\leq 2.
\end{equation}
The following subadditive inequality holds for the function $G$: there exists a constant $C_2$ such that
$$
G(z_1+z_2)\leq C_2 (G(z_1)+G(z_2)) \quad \text{for all} ~~z_1,z_2 \geq 0,~~ z_1+z_2\leq 2,\quad 0\leq \gamma\leq 1.
$$
The last two inequalities imply the statement of the theorem.
\qed

Theorem \ref{tlt} remains valid in the lattice case without the assumption of the boundedness of the potential.
\section{Appendix}
Here we obtain estimates on the solution $p_1$ of problem (\ref{p12}) as $t\gg |x|^2\to \infty$ which provide a rigorous proof of Theorem \ref{thp1}.
\begin{lemma}\label{ogr}
The following estimate holds for $p_1$:
\begin{equation} \label{p1loc11}
|p_{1}(t,x,y)|\leq\frac{C}{t\ln^2 t},~~|x|,|y|\leq 2,~~t>2.
\end{equation}
\end{lemma}
\textbf{Proof.}
Consider the operator $\Delta-q(x)$ in $L^2(R^2)$. This operator is negative, and its spectrum coincides with the semi-axis $(-\infty,0]$, i.e., the resolvent $R^{(1)}_{\lambda}=(\Delta-q(x)-\lambda)^{-1}$ is analytic in $\lambda\in C'=C\setminus (-\infty,0]$. We apply the Laplace transform to (\ref{p12}) and arrive at
\[
p_1=-\int_{a-i\infty}^{a+i\infty}R_{\lambda}^{(1)}(x,y)e^{\lambda t}d\lambda, \quad a>0,~x \neq y,
\]
where $R_{\lambda}^{(1)}(x,y)$ is the kernel of operator $R^{(1)}_{\lambda}$, i.e.,
\begin{equation} \label{rsl}
(\Delta-q(x)-\lambda)R_{\lambda}^{(1)}(x,y)=\delta_y(x).
\end{equation}
We deform the contour of integration in the integral above:
\begin{equation} \label{Ga}
p_1=-\int_{\Gamma} R_{\lambda}^{(1)}(x,y)e^{\lambda t}d\lambda, \quad x \neq y,
\end{equation}
where the contour $\Gamma$ consists of the bisectors of the third and second quadrants of the $\lambda-$plane with the  direction on $\Gamma$ such that Im$\lambda$ increases when a point moves along $\Gamma$. This deformation of the contour is possible since for each fixed $x\neq y$ the function $R_{\lambda}^{(1)}(x,y)$ is analytic in $\lambda\in C'$, decays as $|\text{Im} \lambda |\to \infty$ and is bounded as $\lambda\in C', \lambda \to 0.$ The boundedness will be justified below when the asymptotic behavior of $R_{\lambda}^{(1)}(x,y)$ as $\lambda\to 0$ is established (see (\ref{lto01}), (\ref{lto0})).

Our next goal is to find the asymptotic expansion of $R_{\lambda}^{(1)}(x,y)$ as $\lambda\to 0.$ Let us represent $R_{\lambda}^{(1)}(x,y)$ in the form
\begin{equation} \label{lto01}
R_{\lambda}^{(1)}(x,y)=\frac{\chi(x)}{2\pi}\ln|x-y|+u(\lambda,x,y), \quad |y|\leq 2,
\end{equation}
where $\chi\in C_0^{\infty},~\chi=1$ for $|x|<3,~\chi=0$ for $|x|>4.$ Since $R_{\lambda}^{(1)}(x,y)$ has a logarithmic singularity when $x\to y$, the function $u$ is bounded when $|x|,|y|<3$ and $\lambda>0$ is fixed. We intend to show that
\begin{equation} \label{lto0}
u=u_0(x,y)+\frac{1}{\ln\lambda}u_1(x,y)+\frac{1}{\ln^{2}\lambda}v(\lambda,x,y)), \quad |u_0|,|u_1|,|v|<C,
\end{equation}
when $|x|,|y|\leq 2,~~\lambda \in C',~~|\lambda|<1.$

We put (\ref{lto01}) in (\ref{rsl}) and arrive at
\[
(\Delta-q(x)-\lambda)u=f_1+\lambda f_2,
\]
where
\[
f_1=\frac {1}{\pi}\nabla \chi \cdot \bigtriangledown \ln|x-y|+
\frac {1}{2\pi}(\Delta \chi-q(x)) \ln|x-y|,~~f_2=-\frac{\chi(x)}{2\pi}\ln|x-y|.
\]
Obviously $\|f_1\|_{L^2},~\|f_2\|_{L^2}<C,~|y|\leq 2$, and $f_1=f_2=0$ for $|x|>4$.

In order to describe the behavior of the function $u$ when $|\lambda|<1$, we consider the truncated resolvent
\[
\widehat{R}^{(1)}_{\lambda}=T_2R^{(1)}_{\lambda}T_1:L^2_{com} (R^2)\to L^2_{loc}(R^2), \quad \lambda\in C'=C\setminus (-\infty,0],
\]
of the operator $\Delta-q(x).$ Here $T_1:L^2_{com}(R^2)\to L^2(R^2)$ and $T_2:L^2(R^2)\to L^2_{loc}(R^2)$ are the imbedding operators. Thus, the truncated resolvent is defined on the space of the square integrable  functions with compact supports, and the images are restricted to bounded regions in $R^2.$
The following facts can be found in \cite{vain68, vain75}: the truncated resolvent $\widehat{R}^{(1)}_{\lambda}$ is analytic in $C'$, admits a meromorphic continuation on the Riemannian surface of the function $\ln \lambda$ and has the following asymptotic behavior at the origin
\begin{equation} \label{reso}
\widehat{R}^{(1)}_{\lambda}=A_0+\frac{1}{\ln\lambda}A_1+\frac{1}{\ln^{2}\lambda}B(\lambda), \quad A_0,A_1,B(\lambda):
L^2_{com} (R^2)\to L^2_{loc}(R^2),
\end{equation}
where the operators $A_0,A_1,B(\lambda)$ are bounded and $\|B(\lambda)\|<C$ as $\lambda \to 0,~\lambda \in C'.$ The constant $C$ in the latter estimate depends on the size of the supports of the functions $f$ in the domain of the operator $B$ and the size of the domain in  $R^2$ where the functions $Bf$ are considered.

The validity of (\ref{reso}) needs an explanation. In fact, (\ref{reso}) holds in our case only because $q(x)\geq 0$. In the case of more general potentials $q(x)$ (or more general operators), expansion (\ref{reso}) has a more complicated structure \cite{vain75} involving half-integer powers of $\lambda$ and polynomials of $\ln \lambda$. Formula (\ref{reso}) is an immediate consequence of this more general result in the case when $\|\widehat{R}^{(1)}_{\lambda}\|<C$ as $\lambda \to 0,~\lambda \in C'.$ The uniqueness of the solution of the problem
\begin{equation} \label{uniq}
(\Delta-q(x))w=0,~~x\in R^2; \quad |w|<C, |\nabla w|<C|x|^{-2}~~\text{as} ~ x \to \infty
\end{equation}
implies (see \cite{vain75}) the boundedness of $\|\widehat{R}^{(1)}_{\lambda}\|$ as $\lambda \to 0,~\lambda \in C'$,  and therefore it leads to (\ref{reso}). It remains to note that
\[
0=<(\Delta-q(x))w,w>=-\int_{R^2}(|\nabla w|^2+q(x)|w|^2)dx
\]
for solutions of (\ref{uniq}). Thus $w=0$ if $q(x)\geq 0$ and $q$ is not identically equal to zero. Hence, (\ref{reso}) holds.

We imply the standard a priory estimates (for the Sobolev space $H^2_{loc}(R^2)$) followed by the Sobolev imbedding theorem  and replace the space $L^2_{loc}$ in (\ref{reso}) by the space $C(|x|\leq 2)$ of continuous functions on the disk $|x|\leq 2$. This and the formula $u=\widehat{R}^{(1)}_{\lambda}(f_1+\lambda f_2)$ complete the proof of expansion (\ref{lto0}).

We will also need an estimate on the function $u$ in (\ref{lto01}) when Im$\lambda\to\infty$. Since $u=R_{\lambda}^{(1)}(f_1+\lambda f_2)$ and the norm of the resolvent $R_{\lambda}^{(1)}$ does not exceed the inverse distance from the spectrum, we obtain that
\[
\|u\|_{L^2(R^2)}<C,~~ |y|\leq 2,~~ \lambda \in \Gamma_1=\Gamma \bigcap \{\lambda:|\lambda|>1\}.
\]
This and the standard a priory estimate for elliptic equations imply that $\|u\|_{H^2(R^2)}<C|\lambda|$, where $H^2$ is the Sobolev space. Using  the Sobolev imbedding theorem, we arrive at
\begin{equation} \label{si}
|u|<C|\lambda|,~ |x|,|y|\leq 2,~ \lambda \in \Gamma_1.
\end{equation}

Let us substitute expression (\ref{lto01}) for $R_{\lambda}^{(1)}(x,y)$ in (\ref{Ga}). Since $\int_{\Gamma}e^{\lambda t}d\lambda=0$ for $t>0$, we arrive at
\[
p_1=-\int_{\Gamma} u(\lambda,x,y)e^{\lambda t}d\lambda, ~x \neq y.
\]
Using (\ref{si}), we obtain
\[
p_1=-\int_{\Gamma_2} u(\lambda,x,y)e^{\lambda t}d\lambda+O(e^{-\varepsilon t}), ~~|x|,|y|\leq 2,~~x \neq y,~~t\geq 2,
\]
where $\Gamma_2=\Gamma\setminus\Gamma_1=\Gamma \bigcap \{\lambda:|\lambda|<1\}$ and the estimate of the remainder is uniform in $|x|$ and $|y|$ (one can take $\varepsilon=\sqrt 2 /2$). The integral $\int_{\Gamma_2}e^{\lambda t}d\lambda$ can be evaluated. Since it has order $O(e^{-\varepsilon t}),~t\to\infty,$ expansion (\ref{lto0}) implies
\[
|p_1|\leq C_1|\int_{\Gamma_2}\frac{1}{\ln\lambda} e^{\lambda t}d\lambda|+C_2\int_{\Gamma_2}\frac{1}{|\ln^2\lambda|} e^{\text{Re}\lambda t}d|\lambda|+C_3e^{-\varepsilon t}, ~~|x|,|y|\leq 2,~~x \neq y,~~t\geq 2.
\]
One can replace here the integrals over $\Gamma_2$ by the same integrals over $\Gamma$ plus the terms that decay exponentially as $ t\rightarrow \infty$. Then we make the substitution $\lambda=\mu/t$ in the second integral which leads to the estimate of this integral by $\frac{C}{t\ln^2t},~t>2.$ Thus
\[
|p_1|\leq C_1|\int_{\Gamma}\frac{1}{\ln\lambda} e^{\lambda t}d\lambda|+\frac{C}{t\ln^2t}, ~~|x|,|y|\leq 2,~~x \neq y,~~t\geq 2.
\]
This formula together with (\ref{tt})
prove (\ref{p1loc11}) for $x\neq y$. The latter restriction can be dropped since $p_1$ is continuous when $t>0.$
\qed

The next lemma provides an estimate on $p_1$ when $|x|$ is not bounded.
\begin{lemma}\label{ogr2}
Let $|y|\leq 3/2, |x|\geq 2$ and $t\geq a|x|^{2}\ln|x|$ for some $a>0.$ Then
\begin{equation} \label{p1loc21}
|p_{1}(t,x,y)|\leq\frac{C\ln|x|}{t\ln^2 t},~~C=C(a).
\end{equation}
\end{lemma}

\textbf{Proof.} Consider the following function $v=v(\lambda,x)=\frac{K(\sqrt\lambda |x|)}{\ln \lambda}$, where \[K(\mu)=K_0(\mu)=\frac {\pi i}{2}H^{(1)}_0(i\mu),
~ \mu>0,
\]
is the modified Bessel function ( it is proportional to the Hankel function of the purely imaginary argument). The function $v$ is the exponentially decaying, as $|x|\to \infty$, solution of the problem
\[
(\Delta-\lambda)v=0, ~|x|>2, \quad v|_{|x|=2}=h(\lambda):=\frac{K(2\sqrt\lambda )}{\ln \lambda}, \quad \lambda \in C'=C\setminus (-\infty,0].
\]

Let
\[
\psi(t,x)=\int_\Gamma v(\lambda,x)e^{\lambda t}d\lambda,
\]
where $\Gamma$ is the contour introduced in the proof of the previous lemma. Since $v$ is analytic in $\lambda \in C'$ and decays exponentially when $|\arg \lambda|\leq 3\pi/4, |\lambda|\to\infty$, the integral $\int_\Gamma vd\lambda$ vanishes, and $\psi$ is the solution of the problem
\[
\psi_t=\Delta \psi, ~|x|>2; \quad \psi|_{|x|=2}=\int_\Gamma h(\lambda)e^{\lambda t}d\lambda;
~~\psi|_{t=0}=0.
\]
It will be shown below that the following estimates are valid for the function $\psi$:
\begin{equation} \label{psi1}
\psi|_{|x|=2}=\frac{c_1}{t\ln^2 t}+O(\frac{1}{t\ln^3 t}), \quad t\to \infty,
\end{equation}
\begin{equation} \label{psi2}
|\psi|\leq \frac{C\ln |x|}{t\ln^2 t}, \quad |x|>2, ~~t>a|x|^{2}\ln|x|.
\end{equation}
In particular, (\ref{psi1}) and Lemma \ref{ogr} imply the existence of constants $A$ and $\tau$ such that $A\psi > p_1$ when $|x|=2, t\geq \tau.$

We will also need the solution of the following parabolic problem
\begin{equation} \label{prphi}
\phi_t=\Delta \phi, ~|x|>2; \quad \phi|_{|x|=2}=g(t);
~~\phi|_{t=0}=0,
\end{equation}
where $g(t)=0$ for $t>\tau$, $g(t)=1$ for $t\leq\tau$. We chose a constant $b$ large enough, so that
\[
b>\max_{|x|=2, ~t\leq \tau}(p_1+A\psi).
 \]
Then from the maximum principle it follows that $A\psi+b\phi\geq p_1$ for all $(t,x):t>0,|x|>2$. It will be also shown that estimate (\ref{psi2}) holds for the function $\phi$ when $t>\tau$:
\begin{equation} \label{phi2}
|\phi|\leq \frac{C\ln |x|}{t\ln^2 t}, \quad |x|>2, ~~t>\max(\tau, a|x|^{2}\ln|x|).
\end{equation}
Since $\phi$ is bounded in any bounded region, (\ref{phi2}) implies the same estimate without the restriction $t>\tau$. The latter estimate together with (\ref{psi2}) imply (\ref{p1loc21}). Thus the proof of Lemma \ref{ogr2} will be complete as soon as (\ref{psi1}), (\ref{psi2}), and (\ref{phi2}) are justified.

Let us justify (\ref{psi1}). From the logarithmic behavior of the function $K(\sqrt\lambda)$ at zero and the exponential decay at infinity it follows that
\begin{equation} \label{expK}
K(\sqrt\lambda)=-\ln\sqrt\lambda + \alpha+f(\lambda), \quad |f(\lambda)|<C|\lambda\ln\lambda|, ~\lambda \in \Gamma,
\end{equation}
where $\alpha$ is a real constant.
We took into account here (and below) that $\ln\lambda \neq 0$ when $\lambda \in \Gamma.$ Thus
\[
h(\lambda)=\frac{K(2\sqrt\lambda )}{\ln \lambda}=-1+\frac{\alpha}{\ln\lambda} +q(\lambda), \quad |q(\lambda)|<C|\lambda |, ~\lambda \in \Gamma.
\]
Expansion  (\ref{psi1}) is a consequence of (\ref{tt}) and the following two obvious relations:
\[
\int_\Gamma e^{\lambda t}d\lambda=0~~ \text{for}~~ t>0, \quad |\int_\Gamma q(\lambda) e^{\lambda t}d\lambda|<\frac{C}{t^2}.
\]

Let us justify (\ref{psi2}). Expansion (\ref{expK}) implies
\begin{equation} \label{phi3}
\psi=\int_\Gamma \frac{K(\sqrt\lambda |x|)}{\ln \lambda}e^{\lambda t}d\lambda=\int_\Gamma (-\frac{1}{2}+\frac{\alpha-\ln|x|}{\ln\lambda})e^{\lambda t}d\lambda+\int_\Gamma \frac{f(\lambda |x|^2)}{\ln \lambda }e^{\lambda t}d\lambda.
\end{equation}
Inequality (\ref{psi2}) with arbitrary $t>2$ holds for the first integral in the right-hand side above due to (\ref{tt}). In order to estimate the second integral on the right, we note that
\[
|\frac{f(\lambda |x|^2)}{\ln \lambda }|\leq C|\frac{\lambda |x|^2\ln(\lambda |x|^2)}{\ln \lambda}|\leq
C|\lambda||x|^2+C|\frac{\lambda |x|^2\ln|x|}{\ln \lambda}|.
\]
Thus, the substitution $\lambda \to \mu/t$ implies that the second integral does not exceed
\[
C(\frac{|x|^2}{t^2}+\frac{|x|^2\ln|x|}{t^2\ln t}), \quad |x|\geq 2, t\geq 2.
\]
Hence, (\ref{psi2}) holds if  $|x|\geq 2,~\frac{|x|^2}{t}\leq c\min(\frac {\ln |x|}{\ln^2t},\frac {1}{\ln t})$ with some $c>0.$ It remains to note that the latter restrictions on $x,t$ with some $c=c(a)$ follow from those imposed in the statement of Lemma \ref{ogr2} since the equation $t=a|x|^2\ln|x|$ implies that $\ln t/ \ln |x| \to 2$ as $|x|\to\infty.$ Hence, (\ref{psi2}) holds.

In order to prove the validity of (\ref{phi2}) we solve problem (\ref{prphi}) using the Laplace transform. This leads to
\[
\phi=\frac{b}{2\pi}\int_{-i\infty}^{i\infty} \widehat{\phi}(\lambda,x)e^{\lambda t}d\lambda, \quad \widehat{\phi}(\lambda,x)= \frac{K(\sqrt\lambda |x|)}{K(2\sqrt\lambda) }\frac{1-e^{-\lambda \tau}}{\lambda}.
\]
The contour of integration above can be replaced by $\Gamma$ when $t>\tau:$
$$
\phi=\frac{b}{2\pi}\int_\Gamma \widehat{\phi}(\lambda,x)e^{\lambda t}d\lambda, \quad ~~t>\tau.
$$
Since $K(2\sqrt \lambda)\neq 0$ in $C'$,
expansion (\ref{expK}) leads to the following representation of $\widehat{\phi}(\lambda,x)$ when $\lambda \in \Gamma_2=\Gamma \bigcap \{\lambda:|\lambda|<1\}~:$
\[
\widehat{\phi}=\frac{-\ln(\sqrt\lambda|x|)+\alpha+f(\lambda|x|^2)}{-\ln\sqrt\lambda+\alpha+O(\lambda\ln\lambda)}
(\tau+O(\lambda))=
\tau+\frac{2\tau\ln|x|}{\ln\lambda-2\alpha}+O(\lambda)\ln|x|+O(\frac{f(\lambda|x|^2)}{\ln\lambda}).
\]

The estimate
\[
|\int_{\Gamma_2} O(\frac{f(\lambda|x|^2)}{\ln\lambda})e^{\lambda t}d\lambda|\leq \frac{C\ln|x|}{t\ln^2 t}, \quad |x|\geq 2,~ t\geq\alpha|x|^2\ln|x|,
\]
was proved in the process of evaluating of the second term in the right hand side of (\ref{phi3}). Further,
\[
|\ln|x|\int_{\Gamma_2} O(\lambda)e^{\lambda t}d\lambda|\leq C\ln|x|\int_{\Gamma_2}|\lambda|e^{\text{Re}\lambda t}d|\lambda|\leq C\ln|x|\int_{\Gamma}|\lambda|e^{\text{Re}\lambda t}d|\lambda|=\frac{C\ln|x|}{t^2},
\]
and the integral $\int_{\Gamma_2}e^{\lambda t}d\lambda$ can be evaluated, which implies that
\[
\int_{\Gamma_2}e^{\lambda t}d\lambda=O(e^{-\varepsilon t}),~~t\to\infty.
\]
The last three estimates prove the validity of (\ref{phi2}) for the function
\[
\phi_2:=\frac{1}{2\pi}\int_{\Gamma_2} \widehat{\phi}(\lambda,x)e^{\lambda t}d\lambda
\]
if we take into account that
\[
\int_{\Gamma_2} \frac{1}{\ln\lambda-2\alpha}e^{\lambda t}d\lambda=\int_{\Gamma} \frac{1}{\ln\lambda-2\alpha}e^{\lambda t}d\lambda-\int_{\Gamma_1} \frac{1}{\ln\lambda-2\alpha}e^{\lambda t}d\lambda,~~\Gamma_1=\Gamma \backslash \Gamma_2,
\]
where the following estimates hold for the terms in the right-hand side:
\[
|\int_{\Gamma_1} \frac{1}{\ln\lambda-2\alpha}e^{\lambda t}d\lambda|\leq C \int_{\Gamma_1}e^{\text{Re}\lambda t}d|\lambda|
=O(e^{-\varepsilon t}),~~t\to\infty,
\]
\[
|\int_{\Gamma} \frac{1}{\ln\lambda-2\alpha}e^{\lambda t}d\lambda|\leq \frac{C}{t\ln^2t}, ~~t>2.
\]
The latter inequality can be proved absolutely similarly to (\ref{tt}). Hence, it remains to show that  (\ref{phi2}) holds for the function
\[
\phi_1:=\frac{1}{2\pi}\int_{\Gamma_1} \widehat{\phi}(\lambda,x)e^{\lambda t}d\lambda,~~t>\tau.
\]

We note that $K(2\sqrt\lambda)$ does not vanish on $\Gamma$ and decays exponentially at infinity. This implies that $|\frac{K(\sqrt\lambda |x|)}{K(2\sqrt\lambda) }|\leq C$ when $\lambda \in \Gamma_1,~|x|\geq 2$. Since Re$ \lambda \leq -\sqrt 2/2$ on $\Gamma_1$, it follows that
$$
|\phi_1|\leq C\int_{\Gamma_1}|\frac{1-e^{-\lambda \tau}}{\lambda}e^{\lambda t}d\lambda|<C(\tau)e^{-t\sqrt 2/2}~~t>\tau.
$$
\qed
\begin{lemma}\label{ogr3}
Let $|x|\geq 2$ and $t\geq a|x|^{2}\ln|x|$ for some $a>0.$ Then
\begin{equation} \label{p1loc2}
|p_{1}(t,x,x)|\leq\frac{C\ln^2|x|}{t\ln^2 t},~~C=C(a).
\end{equation}
\end{lemma}

\textbf{Proof.} While $x$ can be an arbitrary point of $R^2$ below, it is always assumed that $|y|\geq 2$ (otherwise the estimate of $p_1$ is provided by Lemma \ref{ogr}). Let $\chi=\chi(t,x)\in C^\infty,~ t\geq 0,$ be a cut-off function such that $\chi=1$ when $|x|,t\leq 1,$ $\chi=0$ when $|x|\geq 3/2$ or $t\geq 2$. We are looking for $p_1(t,x,y)$ with $|y|\geq 2$ in the form
\[
p_1(t,x,y)=p_0(t,x,y)-\chi(t,x)p_0(t,x,0)+z(t,x,y), \quad p_0(t,x,y)=\frac{e^\frac{-|x-y|^2}{4t}}{4\pi t}.
\]
Then $z$ is the solution of the problem
\begin{equation} \label{parz}
z_{t}=\Delta z-q(x)z-f,~~x\in R^2,~~z(0,x,y)=\delta_y(x),
\end{equation}
where
\[
f=q(x)[\chi(t,x)p_0(t,x,0)-p_0(t,x,y)]+2\nabla \chi\cdot\nabla p_0(t,x,0)+(\Delta \chi-\chi_t) p_0(t,x,0)\in C^\infty.
\]

Note that the following estimate holds for the
function $h=\chi(t,x)p_0(t,x,0)-p_0(t,x,y)$:
\[
|h(t,x,x)|\leq \frac{C|x|^2}{t^2 } \quad \text {when} ~~|x|\geq 2, ~t\geq 2.
\]
In particular (see the proof of (\ref{psi2})), from here it follows that
\begin{equation} \label{hh}
|h(t,x,x)|\leq \frac{C(a)\ln|x|}{t\ln^2t } \quad \text {when} ~~|x|\geq 2,~~t\geq a|x|^{2}\ln|x|.
\end{equation}
Hence, it is enough to prove the statement of the lemma for the function $z$ instead of $p_1$.

We need some estimates on the function $f$ in order to estimate $z$. Note that
\begin{equation} \label{fff}
f=q(x)[\frac{e^{-\frac{|x |^2}{4t}}}{4\pi t}-\frac{e^{-\frac{|x-y |^2}{4t}}}{4\pi t}],~~t\geq 2.
\end{equation}
Assuming that $|y|\geq 2$ and taking into account that $|x|\leq 1$ on the support of $q$ and that $f$ is bounded when $|y|\geq 2,~t\leq 2$, we obtain
\[
\int_0^\infty
|f(t,x,y)|dt\leq C+\int_2^\infty|\frac{e^{-\frac{|x |^2}{4t}}}{4\pi t}-\frac{e^{-\frac{|x-y |^2}{4t}}}{4\pi t}|dt
\leq C+\int_2^\infty\frac{1-e^{-\frac{|x |^2}{4t}}}{4\pi t}dt+\int_2^\infty\frac{1-e^{-\frac{|x-y |^2}{4t}}}{4\pi t}dt
\]
\begin{equation} \label{ff}
\leq C_1+\int_2^\infty\frac{1-e^{-\frac{|x-y |^2}{4t}}}{4\pi t}dt=C_1+\int_{\frac{8}{|x-y|^2}}^\infty\frac{1-e^{-\frac{1}{\tau}}}{4\pi \tau}d\tau\leq C_3\ln|y|, \quad |y|\geq 2.
\end{equation}

The next estimate for $f$ is valid when $|y|\geq 2,~~t\geq \frac{a}{2}|y|^{2}\ln|y|$. From (\ref{fff}) it follows that
\[
|f(t,x,y)|\leq \frac{C|y|^2}{t^2},~~|y|\geq 2,~~t\geq 2,
\]
and therefore (see the arguments in the proof of (\ref{psi2}))
\begin{equation} \label{finf}
|f(t,x,y)|\leq \frac{C(a)\ln^2|y|}{t\ln^3t },~~|y|\geq 2,~~t\geq \frac{a}{2}|y|^2\ln|y|.
\end{equation}

We solve problem (\ref{parz}) using the Duhamel principle and arrive at
\[
|z(t,x,y)|\leq \int_0^t\int_{|x_0|<3/2}p_1(s,x,x_0)|f(t-s,x_0,y)|dx_0ds.
\]
Hence
\[
|z(t,x,x)|\leq \int_0^{t/2}\int_{|x_0|<3/2}p_1(s,x,x_0)|f(t-s,x_0,x)|dx_0ds
\]
\begin{equation} \label{duam}
+
\int_{t/2}^t\int_{|x_0|<3/2}p_1(s,x,x_0)|f(t-s,x_0,x)|dx_0ds, \quad |x|\geq 2.
\end{equation}

 Function $p_1(s,x,x_0)$ can be estimated through $p_0$ for all values of $s$, and therefore
\[
|p_1(s,x,x_0)|<\frac{C}{1+s} ~~\text {when}~~|x-x_0|>1/2.
\]
This and (\ref{finf}) imply that the first term in the right-hand side of (\ref{duam}) does not exceed $\frac{C(a)\ln^2|y|}{t\ln^2t }$ when $|x|\geq 2,~t\geq \frac{a}{2}|y|^2\ln|y|.$ The same estimate is valid for the second term due to (\ref{ff}) and (\ref{p1loc21}) (where one also needs to keep in mind that $p_1(s,x,y)=p_1(s,y,x)$). Hence (\ref{p1loc2}) is proved for $z$.
\qed

\end{document}